\documentclass[a4paper,11pt,bookmarks=true]{article}
\pdfoutput=1
\usepackage{jheppub}
\usepackage{graphicx}
\usepackage[colorlinks=true,linkcolor=blue,citecolor=blue, urlcolor=blue]{hyperref}   
\usepackage{tikz,calc}
\usetikzlibrary{shapes,patterns,arrows,positioning,automata}
\usepackage{pgfplots}
\def\Eq#1{Eq.~(\ref{#1})}
\def\Eqs#1{Eqs.~(\ref{#1})}
\def\eq#1{(\ref{#1})}

\def\app#1{Appendix~\ref{#1}}
\def\Fig#1{Fig.~\ref{#1}}
\def\Sect#1{Section~\ref{#1}}

\usepackage{bm}
\def\p{\mathbf{p}}
\def\phip{\phi_{\mathbf{p}}}
\def\x{\mathbf{x}}
\def\xp{\mathbf{x}_\perp}
\def\txp{\tilde{\mathbf{x}}_\perp}
\def\phix{\phi_{\mathbf{x}}}
\def\phitx{\phi_{\tilde{\mathbf{x}}}}
\def\phiDtx{\phi_{\Delta\tilde{\mathbf{x}}}}
\def\k{\mathbf{k}}
\def\phik{\phi_{\mathbf{k}}}
\def\pp{\mathbf{p'}}
\def\phipp{\phi_{\mathbf{p'}}}
\def\kp{\mathbf{k'}}
\def\phikp{\phi_{\mathbf{k'}}}

\usepackage{multirow}
\usepackage{dcolumn}

\begin{document}
\preprint{CERN-TH-2021-056, LU-TP 21-12}

\title{Collective flow in single-hit QCD kinetic theory}
\author[a,b]{Aleksi Kurkela,}
\emailAdd{aleksi.kurkela@cern.ch}
\affiliation[a]{Theoretical Physics Department, CERN, CH-1211 Geneva 23, Switzerland}
 \affiliation[b]{Faculty of Science and Technology, University of Stavanger,  4036 Stavanger, Norway}
\author[a]{Aleksas Mazeliauskas,}
\emailAdd{aleksas.mazeliauskas@cern.ch}
\author[c]{Robin Törnkvist}
\emailAdd{robin.tornkvist@thep.lu.se}
\affiliation[c]{Department of Astronomy and Theoretical Physics, Lund University, SE-221 00 Lund, Sweden}

\date{\today}

\abstract{
Motivated by recent interest in collectivity in small systems, we  calculate the harmonic flow response to initial geometry deformations within weakly coupled QCD kinetic theory using the first correction to the free-streaming background.  We derive a parametric scaling formula that relates harmonic flow in systems of different sizes and different generic initial gluon distributions. We comment on similarities and differences between the full QCD effective kinetic theory and the toy models used previously. Finally we calculate the centrality dependence of the integrated elliptic flow $v_2$ in oxygen-oxygen, proton-lead and proton-proton collision systems.
}

\maketitle
\section{Introduction}

The physical picture implemented in event generators used in phenomenological modelling of soft particle production in proton-proton collisions \cite{Sjostrand:2014zea,Bellm:2015jjp} is maximally different from that implemented in the descriptions of nucleus-nucleus collisions~\cite{Heinz:2013th,Gale:2013da}. 
In the former, partons fragment but free-stream to the detector after being created in an initial hard scattering without undergoing secondary rescatterings with the other partons. In the latter, reinteractions are so strong that a nearly ideally hydrodynamized fluid is created. 
The recent experimental characterization of the smooth onset of collectivity, \emph{i.e.}, the onset of qualitative features characteristic to the fluid-dynamic picture and absent in the free-streaming picture, challenges this dichotomy of two disconnected pictures~\cite{Citron:2018lsq}. Prime examples of the signs of collectivity include the strangeness enhancement observed in proton-proton ($pp$), proton-nucleus ($p$A), and in nucleus-nucleus (AA) collisions \cite{ALICE:2017jyt} as well as the formation of multi-particle collective flow, measured by the azimuthal harmonics $v_n$ and their cumulants~\cite{Abelev:2014mda,Khachatryan:2015waa,Sirunyan:2017uyl}. 
These observations offer an inroad to study how the ideal fluid is built up from fundamental interactions of elementary particles and how hydrodynamization takes place as a function of the system size \cite{Citron:2018lsq}. 

There have been multiple attempts to describe collectivity in small systems extending the models of $pp$ and AA collisions~\cite{Adolfsson:2020dhm,Nagle:2018nvi}. 
However, 
in order to fully exploit the experimental progress, the experimental data needs to be confronted with models that encompass the both extremes and are able to dynamically describe the process of hydrodynamization. One such a model is the QCD effective kinetic theory (EKT)~\cite{Arnold:2002zm} that gives a leading-order accurate description of the matter created in the asymptotic limit of infinite collision energy $\sqrt{s}\rightarrow \infty$. It builds on a picture arising from perturbative QCD in which partons undergoing elastic scattering and radiate medium-induced collinear radiation. This effective theory has been extensively employed to study how non-abelian gauge theories thermalize and how the hydrodynamization takes place in heavy-ion collisions in the weak coupling limit~\cite{York:2014wja, Kurkela:2015qoa, Kurkela:2018wud, Kurkela:2018xxd, Almaalol:2020rnu, Du:2020zqg, Du:2020dvp, Kurkela:2019set,Mazeliauskas:2018yef} (see also~\cite{Schlichting:2019abc,Berges:2020fwq}). It has also been used to study how hard partons are hydrodynamized in the QCD medium leading to jet quenching~\cite{Kurkela:2014tla, Schlichting:2020lef} (see also \cite{DEramo:2018eoy,Iancu:2015uja,Blaizot:2013hx,Ghiglieri:2015zma}). Here, we will take the first steps to employ the EKT
to study how hydrodynamization takes place as a function of system size and specifically how signals of collectivity arises in the smallest systems.  

In the case of sufficiently small systems, the system stays far from equilibrium throughout the evolution and the deviation from collisionless expansion can be treated as a perturbation to the free-streaming background. This single-hit approximation has been studied for multiple simplified models of QCD medium capturing different aspects of the full effective kinetic theory, including the isotropization time approximation (ITA) \cite{Kurkela:2018ygx,Kurkela:2018qeb,Kurkela:2019kip,Kurkela:2020wwb} and other generic models of elastic scattering \cite{Heiselberg:1998es,Kolb:2000fha,Borghini:2010hy,Roch:2020zdl, Borghini:2018xum,Borghini:2010hy,Romatschke:2018wgi}. 
These works have demonstrated that the first scatterings have the largest effect in formation of the flow harmonics and that already the single-hit approximation leads to sizeable elliptic flow $v_2$. This qualitative observation suggests that collectivity in a form of azimuthal flow is a signal of final state interactions rather than of full hydrodynamization. In ref.~\cite{Kurkela:2020wwb}, the systematics of how different flow harmonics arise in kinetic theories have been characterized and contrasted with how they arise in relativistic fluid dynamics. 

The QCD effective kinetic theory has more structure than the simple ITA kinetic theory. In ITA, the response of the flow coefficients to initial geometry depends only on a single parameter, opacity $\hat \gamma$, that describes the system size in units of the mean free path \cite{Kurkela:2018ygx}. The situation in EKT is somewhat more complicated because the collision kernel depends non-trivially on the interplay of  classical and Bose-enhanced parts of the scattering terms as well as on the in-medium screening that regulates the soft divergence of the elastic scattering. Hence, in addition to depending on the system size, the flow is sensitive to the occupancy determining the strength of Bose enhancement as well as the coupling constant, determining the amount of Debye screening. Here, we will fully characterize the energy weighted linear flow response coefficient $v_2/\varepsilon_2$ in terms of these variables. 

We note that while EKT gives a leading order description of time-evolution of large and isotropic systems, in the current work we are pushing it beyond its strict applicability. Hence, our results are not fully leading-order accurate but represent rather a model calculation that is based on our current best understanding of perturbation theory.  
The manuscript is organized as follows. 
In \Sect{sec:methods} we summarize the model setup and discuss the computation of the linear response function $v_n/\varepsilon_n$ using the co-moving coordinate system.  In \Sect{sec:results} we discuss the conformal scaling of EKT results and make comparisons to ITA. Finally, we present an exploratory study of elliptic flow signal in oxygen-oxygen (OO), proton-lead ($p$Pb) and proton-proton ($pp$) collisions using single hit EKT.

\section{Methods}
\label{sec:methods}

\subsection{Single hit approximation}
We consider the leading order QCD effective kinetic theory formulation by Arnold, Moore and Yaffe~\cite{Arnold:2002zm}. In a boost invariant system the color and polarization averaged distribution $f$ is governed by the Boltzmann equation \cite{Baym:1984np}
    \begin{equation}
        \left[\frac{\partial}{\partial \tau}+\bm{v}_\perp \cdot \frac{\partial}{\partial \xp} - 
        \frac{p_z}{\tau}\frac{\partial}{\partial {p_z}}\right]f(\tau,{\bf x_\perp};{\bf  p_\perp}, p_z)
        = - C [f].\label{eq:Boltzmann}
        \end{equation}
The collision kernel $C [f] = C_{2\leftrightarrow 2}[f]+ C_{1\leftrightarrow 2} [f]$ is a sum of a leading-order elastic     $2\leftrightarrow 2$ scattering kernel containing the physics of in-medium screening and an effective medium-induced collinear $1\leftrightarrow 2$ radiation kernel. For the explicit discussion of the implementation we refer to previous papers~\cite{Kurkela:2015qoa,Kurkela:2018oqw}. In the following we will restrict ourselves only to a purely gluonic plasma.  
        
We are interested in small, dilute systems, where partons experience only a few scatterings. Therefore it is useful to first consider the solution to
 the collisionless Boltzmann equation $f^{(0)}$. The free-streaming solution 
 can be written entirely in terms of the initial distribution given at time $\tau_0$ using a co-moving coordinate system \cite{Kurkela:2018qeb}
 $$
 f^{(0)}(\tau,\xp ; \p_\perp,p_z) =f^{(0)}(\tau_0,\txp; \p_\perp, \tilde p_z),
 $$
 where the longitudinal momentum is simply the rescaled $\tilde p_z=p_z  \frac{\tau}{\tau_0}$, while the transverse co-moving coordinate is
    \begin{align}
        \label{eq:xtilde}
        \txp &=\xp-\frac{\bm{p}_\perp}{p_\perp^2} \tau \left( \sqrt{p_\perp^2+p_z^2}- \frac{\tau_0}{\tau}\sqrt{p_\perp^2+p_z^2 \frac{\tau^2}{\tau_0^2}} \right).
    \end{align}
 
    In co-moving coordinates, defining the co-moving distribution function $\tilde f(\tau, \tilde \x_\perp; \p_\perp, \tilde p_z) = f(\tau, \x_\perp; \p_\perp, p_z)$, the Boltzmann equation reduces to an ordinary differential equation in time $\tau$,
    \begin{equation}
        \partial_\tau \tilde f(\tau,\txp; \p_\perp, \tilde p_z) = - \tilde C[\tilde f] (\tau,\txp; \p_\perp, \tilde p_z).
    \end{equation}
   This form is convenient for linearizing the solution \emph{in the number of scatterings} $\tilde f= \tilde f^{(0)}+\tilde f^{(1)}+\ldots$, where the  background and the single-scattering distributions evolve according to
    \begin{align}
    \begin{split}
        \partial_\tau \tilde f^{(0)} &= 0,\\
        \partial_\tau \tilde f^{(1)} &= - \tilde C[ \tilde f^{(0)}].
        \end{split}\label{eq:f1comoving}
    \end{align}
    This is the so-called  single hit approximation and the first correction to the distribution can be obtained directly by integration
    \begin{equation}
        \label{eq:f1sol}
       \tilde{f}^{(1)}(\tau
       )=- \int_{\tau_0}^\tau d\tau'  \tilde C[\tilde f^{(0)}](\tau'
       ),
    \end{equation}
    where the initial condition for the perturbation is $\tilde f^{(1)}(\tau_0)=0$ by definition.
    
    \subsection{Energy flow}
   We will calculate the elliptic flow generated in the single-hit approximation in QCD kinetic theory. The elliptic flow $v_2$ is customarily defined by the azimuthal deformation of the \emph{particle number}. The translation of partonic degrees of freedom to observed hadronic ones is sensitive to the details of the hadronization procedure as there is no conservation of particle number between the partonic and hadronic degrees of freedom of QCD. We instead consider the soft and collinear safe \emph{transverse energy flow} which is insensitive to the hadronization effects. The transverse energy flow per unit rapidity reads
    \begin{equation}
        \label{eq:energyflow}
        \frac{dE_\perp}{d\eta d\phip} = \tau \int d^2\xp\int \frac{p_\perp dp_\perp dp_z}{(2\pi)^3} p_\perp f(\tau,\xp; \p_\perp, p_z)
    \end{equation}
    as a function of the momentum-space azimuthal angle $\phi_\p$, where  $\p_\perp = p_\perp(\cos(\phi_\p),\sin(\phi_\p))$. The energy weighted flow coefficients $v_n$ here are defined as Fourier components of the transverse energy flow
        \begin{equation}
        \label{eq:energyvn}
        \frac{dE_\perp}{d\eta d\phip} = \frac{dE_\perp}{2\pi d\eta }\bigg(1 + 2\sum_{n=1}  v_n \cos[n(\phip-\psi_n)]\bigg),
    \end{equation}
    where $\psi_n$ defines the azimuthal orientation of $n$th harmonic flow. 

Inserting the single hit expansion  $f\approx f^{(0)}+f^{(1)}$ in \Eq{eq:energyflow}
 and noting that the symmetric free-streaming background has vanishing harmonics (assuming no initial momentum-space anisotropy), we get 
    \begin{align}
 &       \frac{dE_\perp}{d\eta d\phi_{p}}  -\frac{dE_\perp}{2\pi d\eta}\bigg\rvert_{f^{(0)}} =\tau \int d^2\xp\int \frac{p_\perp dp_\perp dp_z}{(2\pi)^3}p_\perp f^{(1)}\nonumber\\
     &=\tau \frac{\tau_0}{\tau}\int \frac{p_\perp dp_\perp d\tilde{p}_z}{(2\pi)^3}\int d^2\txp p_\perp \tilde{f}^{(1)},
    \end{align}
    where we changed to co-moving $\tilde{\x}_\perp, \tilde{p}_z$ coordinates and gained a $\tau_0/\tau$ factor as a Jacobian.
    Now we can substitute the single-hit approximation, \Eq{eq:f1sol}, and revert to lab-coordinates.
    \begin{align}
     & =-\int_{\tau_0}^\tau d\tau' \tau_0\int \frac{p_\perp dp_\perp d\tilde{p}_z}{(2\pi)^3}\int d^2\txp p_\perp  \tilde C[ \tilde f^{(0)}]\\
     & =- \int_{\tau_0}^\tau d\tau' \tau '\int d^2\x_\perp\int \frac{p_\perp dp_\perp dp_z}{(2\pi)^3} p_\perp C\left[f^{(0)} \right]. \label{eq:energyflow_perturbed}
    \end{align}
 In leading order QCD the collision kernel consists of two terms: $C= C_{2\leftrightarrow 2}+ C_{1\leftrightarrow 2}$. However the collinear $1 \leftrightarrow 2$ splitting does not change the particle orientation and only shifts particles in energy.
 Therefore in the single hit approximation only $2\leftrightarrow 2$ scatterings will contribute to the soft and collinearly safe transverse energy weighted flow of \Eq{eq:energyflow_perturbed} and momentum integrated $v_n$ is given by
 \begin{align}\label{eq:vnfull}
        v_n    =- \int_{\tau_0}^\tau d\tau' \tau'&\!\!\int \frac{d^2\xp}{2\pi}
        \int\! \frac{d^3\p}{(2\pi)^3} \nonumber\\ \times& \frac{p_\perp \cos[n(\phip-\psi_n)]C_{2\leftrightarrow2}\left[f^{(0)}\right]}{dE_\perp/(2\pi d\eta)}.
  \end{align}

    \subsection{Linearized perturbation}
    
    We further simplify the problem by considering small azimuthal perturbations $\delta f$ of an otherwise symmetric background $\bar f$,
        \begin{equation}
        f^{(0)}(\tau_0,\xp; \p_\perp, p_z) = \bar{f}(\tau_0, x_\perp; \p,p_z)  + \delta f(\tau_0,  x_\perp, \phi_\x; \p,\xp),
    \end{equation}
    with the coordinate-space azimuthal angle $\phi_\x$, where $\x_\perp = x_\perp(\cos(\phi_\x),\sin(\phi_\x))$.
   Then the collision kernel can be also linearized
       \begin{equation}
        C_{2\leftrightarrow 2}\left[f^{(0)}\right]\approx C_{2\leftrightarrow 2}\left[\bar{f}\right]+ \delta C_{2\leftrightarrow 2}\left[\bar{f},\delta f\right].
    \end{equation}
At initial time the background distribution $\bar{f}$ is azimuthally symmetric both in momentum and coordinate space, therefore scatterings of the background term
$C_{2\leftrightarrow 2}\left[\bar{f}\right]$ will not contribute to anisotropy generation. Then, to linear order in perturbations, the momentum integrated $v_n$ is 
 \begin{align}
        v_n    =- \int_{\tau_0}^\tau d\tau' \tau '\int rdr D_n(\tau',r)
        \label{eq:vn}
  \end{align}
where we defined the partial integral $D_n$ as
\begin{equation}
    \label{eq:vnDn}
    D_n=\int\!\frac{d\phix}{2\pi} \!\int\! \frac{d^3\p}{(2\pi)^3}  \frac{p_\perp \cos[n(\phip-\psi_n)]\delta C_{2\leftrightarrow 2}\left[\bar{f},\delta f\right]}{\left.dE_\perp/(2\pi d\eta)\right|_{\bar{f}}}.
\end{equation}
The linearized collision kernel  $\delta C_{2\leftrightarrow 2}\left[\bar{f},\delta f\right]$ 
can be obtained by straightforward linearization of the
collision matrix $|\mathcal M|^2$ (via $m_g^2$) and the phase-space terms. For simple single harmonic perturbations of the background discussed below, the $\phi_\x$ integral in \Eq{eq:vnDn} can be performed explicitly. Then $D_n(\tau,r)$ can be calculated by Monte-Carlo sampling of the multidimensional phase-space integral on a discrete $\tau-r$ grid. The integrated $v_n$ is then found by numerically integrating $D_n(\tau,r)$. We refer to \app{sec:Dn} for further details.

\subsection{Elastic scattering kernel and screening mass}

    The elastic collision kernel is given by the multi-dimensional phase-space integral~\cite{Arnold:2002zm}
\begin{align}
\label{2to2}
 {C}_{2\leftrightarrow 2}[f](\p)&= \frac{1}{4|\p| \nu_g}\int \frac{d^3 
 \k}{2k (2\pi)^3}\frac{d^3 \p'}{2p' (2\pi)^3}\frac{d^3 \k'}{2k' 
 (2\pi)^3}|\mathcal{M}(\p,\k;\p',\k')|^2 (2\pi)^4 \delta^{(4)}(P^\mu+K^\mu-P'^\mu-K'^\mu) 
 \nonumber \\
 &\times \big\{ f_\p f_\k[1+ f_{\p'}][1+ f_{\k'}]-f_{\p'}f_{\k'}[1+ f_{\p}][1+ 
 f_{\k}] \big\},
\end{align}
  where the leading order scattering-amplitude-square is given in Mandelstam variables $s,t,u$ and the 't Hooft coupling constant $\lambda=N_c g^2$ as
\begin{align}
    \label{eq:matrixelement}
    |\mathcal{M}|^2/\nu_g = 2 \lambda^2 \left( 9 + \frac{(s-t)^2}{u^2}+ 
    \frac{(u-s)^2}{t^2}+ \frac{(t-u)^2}{s^2}\right).
\end{align}
 Small angle scattering then exchange momentum $q=|\p-\p'|\to 0$ and the $t$-channel (and similarly $u$-channel) suffers from Coloumb
 divergence. Medium screening effects have to be taken into account. This can be achieved by introducing a regularized $t_\text{reg}$ in the denominator of \Eq{eq:matrixelement} (identically for $u_\text{reg}$),
  \begin{equation}\label{eq:treg}
t \rightarrow t_\text{reg}=  \frac{t(q^2+\frac{e^{5/3}}{4} m_g^2)}{q^2}.
  \end{equation}
  Here  $m_g^2$ is the gluon 
  screening or effective mass
  \begin{align}
    \label{eq:m2}
    m^2_g &= 2 \lambda \int \frac{d^3 p}{(2\pi)^3} \frac{f_\p}{|\p|}.
  \end{align}
  The constant  $e^{5/3}/4$ in \Eq{eq:treg} is chosen to reproduce gluon drag and diffusion in full HTL treatment near equilibrium~\cite{York:2014wja}. We note that the effective mass is Lorentz invariant, which follows from the fact that $d^3p/|\p|$ is a Lorentz invariant integration measure and the phase space density is a  Lorentz scalar~\cite{Treumann:2011zb}. Therefore we can perform the integral in \Eq{eq:m2} in the lab frame.

\subsection{Initial conditions}
\label{subsection:initialconditions}
\subsubsection{Background}
    For initial conditions of the spatially azimuthally symmetric background at an initial time $\tau_0$ we consider the following ansatz designed to capture certain properties of the gluon distribution in a gluon-saturation, or Color-Glass-Condensate (CGC), -based calculation~\cite{Kurkela:2015qoa,Lappi:2011ju},
    \begin{align}
        \label{eq:kzdistribution}
        \bar{f} 
        & = \frac{A}{p_\xi}e^{-\frac{2p_\xi^2}{3}},\quad
        p_\xi \equiv \frac{\sqrt{p_\perp^2+\xi^2p_z^2}}{Q(\xp)},
    \end{align}
    where $A$ controls the magnitude of the occupation. In CGC calculations in the weak-coupling limit $A \propto 1/\lambda \gg 1$. The parameter
    $\xi$ controls the longitudinal momentum asymmetry, which is assumed to be large in CGC-type initial conditions. 
    Here, 
   $Q(\x_\perp)$ is the characteristic energy scale, which can be related to the average transverse momentum
   \begin{equation}
     \left.\left<p_\perp^2\right>\right|_{\tau=\tau_0, \xp}=Q^2(\xp)  .
   \end{equation}
   For the transverse $Q$ profile we take a simple Gaussian of width $\sqrt{2}R_0$,
    \begin{equation}
        Q(\xp) = Q_0 e^{- \frac{|\xp|^2}{4R_0^2}}.
    \end{equation}
Defining a short-hand notation $\hat{A} =\frac{A\tau_0}{\xi R_0}$, the initial (lab-frame) energy density per rapidity (for $\xi\gg 1$) is
\begin{equation}\label{eq:etau0}
\tau_0 e(\tau_0, \xp)\approx \frac{9\nu_g }{64\sqrt{6\pi}} \hat{A} R_0Q_0^4 e^{- \frac{|\xp|^2}{R_0^2}},
\end{equation}
while the transverse energy integral (conserved by free-streaming) is
\begin{align}
    \left.\frac{dE_\perp}{2\pi d\eta}\right|_{\bar f} =\frac{3\sqrt{6\pi}\nu_g}{128} \hat{A}R_0^3Q_0^4  .
    \label{eq:dEperp}
\end{align}
Here $\nu_g$ is the number of gluon degrees of freedom ($\nu_g=16$ for $N_c=3$).

The elastic scattering rate depends on the gluon screening mass, \Eq{eq:m2}. For an anisotropic distribution $\xi \gg 1$ and at the initial time $\tau_0$, it can be written as
\begin{align}
   \tau_0 m_g^2(\tau_0, \xp) &
    \approx \lambda \sqrt{\frac{3 }{32\pi}} \hat{A} Q_0^2 R_0 e^{- \frac{|\xp|^2}{2R_0^2}} .
    \label{eq:mg}
\end{align}

To study the sensitivity of our results to the precise form of the distribution function, we will also consider a deformed Bose-Einstein distribution
\begin{align}
    \label{eq:thermaldistribution}
    \bar{f} = \frac{1}{e^{p_\xi}-1},\quad  p_\xi \equiv \frac{\sqrt{p_\perp^2+\xi^2p_z^2}}{ Q(\xp)}\sqrt{\frac{8 \zeta (5)}{\zeta (3)}},
\end{align}
where $\sqrt{{8 \zeta (5)}/{\zeta (3)}}\approx 2.6$ preserves the equality between $Q$ and the transverse momentum 
$\left.\left<p_\perp^2\right>\right|_{\tau=\tau_0, \xp}=Q^2(\xp)$ at the initial time.

\subsubsection{Perturbation}
For the linearized azimuthal perturbation we take
\begin{equation}
    \label{eq:deltaf}
  \delta f =   \epsilon \bar{f}\frac{x_\perp^n}{R^n_0}\cos(n \phix),
\end{equation}
where $\epsilon \ll 1$ is a number characterising the size of the perturbation (and which can be scaled out from calculations in linearized equations). Note that such a choice of perturbation in \Eq{eq:deltaf} implies that, by symmetry,  $\psi_n=0$ in \Eqs{eq:energyvn} and \eq{eq:vnDn} for the orientation of the flow harmonics.

We will mainly present results  for the elliptic deformation $n=2$.
For such a perturbation we can define the eccentricity with respect to the background energy density
\begin{equation}
    \varepsilon_2 \equiv-
    \frac{\int d^2 \xp x_\perp^2 \cos(2\phix) e(\xp)}{\int d^2\xp x_\perp^2 e(\xp)}=-\epsilon,
\end{equation}
which has an opposite sign to $\epsilon$. As geometry anisotropy is transformed to momentum anisotropy with a sign change, such definition of geometry eccentricty keeps the response coefficient $v_2/\varepsilon_2$ positive.
Similarly we can define $n=3$ eccentricity as
\begin{equation}
    \varepsilon_3 \equiv-
    \frac{\int d^2 \xp x_\perp^3 \cos(3\phix) e(\xp)}{\int d^2\xp x_\perp^3 e(\xp)}=- \frac{4}{\sqrt{\pi}}\epsilon.
\end{equation}

\section{Results}
\label{sec:results}

\subsection{Conformal scaling of the solutions}
\label{sec:conformalscaling}

The EKT of \Eq{eq:Boltzmann} possesses conformal symmetry which connects the flow from simulations performed with different parameters. The different parameters characterizing the initial condition are the spatial extent ($R_0$), mean transverse momentum ($Q_0$), and initial occupancy ($A$). In addition, the initial condition depends on the initial anisotropy ($\xi$) at the initialization time ($\tau_0$). We note that the kinetic theory itself depends on the coupling $\lambda$, both directly setting the overall rate of the collisions as well as determining the ratio $m_g/Q_0$. Lastly, to linear order in perturbations the flow response is linearly proportional to initial geometric deformations characterized by eccentricities $\varepsilon_n$. We first discuss how simulations with different initial conditions are connected to each other through conformal scaling followed by a discussion about how varying the coupling of the kinetic theory affects the flow variables. 

The conformal scaling and the lack of intrinsic scales allows to trivially relate all simulations with different spatial extents and initial transverse momentum magnitude. This can be seen by scaling all coordinate and momentum space variables in \Eqs{eq:vn} and \eq{eq:vnDn} by appropriate powers of $R_0$ and $Q_0$ respectively, i.e.,\ $\hat \tau = \tau/R_0$ and $\hat r = r /R_0$, so that
\begin{align}\label{eq:hatDn2}
        v_n    =- R_0 Q_0\int_{\tau_0/R_0}^{\infty} d\hat{\tau}' \hat{\tau} '\int \hat{r}d\hat{r} \hat{D}_n(\hat \tau',\hat r),
  \end{align}
where the rescaled $\hat{D}_n$ is
\begin{equation}\label{eq:hatDn}
    \hat{D}_n=\int \frac{d\phix}{2\pi} \int \frac{d^3\hat{\p}}{(2\pi)^3}  \frac{\hat{p}_\perp \cos(n\phip)\delta \hat C_{2\leftrightarrow 2}\left[\bar{f},\delta f\right]}{\left.d \hat E_\perp/(2\pi d\eta)\right|_{\bar{f}}}.
\end{equation}
Here, $\hat E_\perp = E_\perp/R_0^3Q_0^4$ is the dimensionless transverse energy. 
Recall that the collision kernel has units of momentum and $\delta \hat C = \delta C/ Q_0$ has absorbed one power of $Q_0$.  After rescaling the dimensionful quantities, $\hat v_n = v_n/(R_0 Q_0)$ depends only on dimensionless parameters, $A$, $\xi$ and $\hat \tau_0 = \tau_0/R_0$.

If we start with the initial particle distribution \emph{isotropic} in the transverse momentum space, but \emph{anisotropic} in coordinate space, most of the final $v_n$ will be generated at times $\tau\sim R_0$ when particles from different density regions will rescatter. In particular, we do not expect flow generation at the initial time $\tau_0 \ll R_0$ and we would therefore like to take the $\hat \tau_0 \rightarrow 0$ limit. At early times $\hat \tau \ll 1$ when the transverse expansion may be neglected, the main effect of the free-streaming expansion is the linear increase of longitudinal anisotropy with time. Therefore a well-behaved limit $\tau_0 \rightarrow 0$ can be taken keeping $\hat \tau_0/\xi$ fixed. 

Furthermore, we note that if the momentum space distribution is anisotropic enough, $\xi \gg 1$, it essentially behaves as a delta function in $p_z$, $f \propto \delta (p_z)$. In this case the dependence in the overall occupancy of the initial distribution, $A$, can be absorbed to either initial time $\hat \tau_0$ or equivalently to initial anisotropy $\xi$.  That is, if there are two initial conditions with a fixed $A/\xi$ but differing $A$ and $\xi$, the one with the larger $A$ will have a larger occupancy but less phase space occupied so that both distributions have the same number of particles and other integral moments. Therefore, keeping the parameters of the theory fixed (i.e., $\lambda$) but varying the initial conditions, the rescaled flow variable $\hat v_n$ depends only on the scaling variable
\begin{equation}
   \hat A \equiv \frac{ A\tau_0}{\xi R_0},
   \label{eq:ICscaling}
\end{equation}
assuming that $\tau_0 \ll R_0$ and $\xi \gg 1$.

Here we discuss the relation of the scaling variable $\hat{A}$ to initial gluon multiplicity for CGC-type initial conditions. In the saturation framework, the initial gluon multiplicity per rapidity and unit transverse area is given by~\cite{Mueller:1999fp,Kovchegov:2000hz} 
\begin{equation}\label{eq:CGCNg}
    \frac{dN_g}{d^2\mathbf{x}_\perp d\eta} = c \frac{\nu_g Q_0^2}{2\pi \lambda},
\end{equation}
where $c\approx \mathcal{O}(1)$  is gluon liberation coefficient~\cite{Lappi:2011ju}. 
On the other hand, the total particle number integral in our parametrization scales as
\begin{equation}
 \frac{d N_g}{d\eta} \propto \hat{A} R_0^3     Q_0^3.
\end{equation}
Therefore $\hat{A}$ has the following dependence on initial gluon multiplicity for CGC-type initial conditions
\begin{equation}
\hat A \propto \frac{1/\lambda}{R_0Q_0}\propto \lambda^{-3/2}\left(\frac{d N_g}{d\eta}\right)^{-1/2}.\label{eq:CGCscaling}
\end{equation}
We note that perturbatively $1/\lambda \propto \log \left(Q_0/\Lambda\right)$, where $\Lambda$ is the non-perturbative QCD scale. For a system composed of nucleons $R_0>1/\Lambda$ and in the perturbative $Q_0\to \infty$ limit we therefore get that $\hat{A}\lesssim 1$.

Now we consider the scaling of the numerator in \Eq{eq:hatDn}.
The loss and gain terms in the collision integral consist of a classical piece, which is quadratic in phase-space distributions and a cubic Bose-enhanced piece. These two terms scale as $A^2$ and $A^3$, respectively. As the dependence on the initial conditions arises only through scaling variables, the dependence on $A$ implies that the two terms are proportional to $\hat{A}^2$ and $\hat{A}^3$.  
Cancelling one power of $\hat{A}$ in the denominator, we can express the flow $v_n$ as a sum of classical and Bose-enhanced terms:
\begin{align}
    &v_n =  v_n^\text{cl.}+v_n^\text{b.e.} \\
    & \approx R_0 Q_0 \varepsilon_n \left( \frac{ A \tau_0}{\xi R_0} \right) \tilde v_n^\text{cl.}+ R_0 Q_0 \varepsilon_n \left( \frac{ A \tau_0}{\xi R_0} \right)^2 \tilde v_n^\text{b.e.}.
 \label{eq:v2scaling:first}
\end{align}

The above \Eq{eq:v2scaling:first} describes how the flow variables change when the initial conditions are varied but the coupling of the kinetic theory itself is kept fixed. Now we will finally discuss how different simulations with different values of $\lambda$ are connected to each other. First we note that the matrix element is explicitly proportional to $\lambda^2$ so that $\tilde v_2  \equiv \lambda^2 \hat v_2 \sim  \lambda^2$. In addition to this explicit dependence, the in-medium screening regulating the matrix element depends on the coupling: a more strongly coupled medium leads to more screening and hence less small-angle scattering. This dependence on $m_g^2/Q_0^2$ is genuinely non-trivial and simulations with different $m_g^2/Q_0^2$ cannot be related to each other through a simple scaling. Therefore both $\hat v_n^\text{cl.}$ and  $\hat v_n^\text{b.e.}$ are non-trivial functions of this ratio. 
As the dependence on the initial conditions must come in the specific combination set by the scaling variable of \Eq{eq:ICscaling}, the dependence on $m_g^2/Q_0^2$ can only enter through
\begin{align}
    \hat m_g^2 = \frac{m_g^2(\tau_0,r=0)}{Q_0^2} \frac{\tau_0 }{R_0},
\end{align}
where we used the fact that 
$m_g^2(\tau_0,r=0) \sim \lambda A Q_0^2/\xi$, see \Eq{eq:mg}.
We finally note that the two scaling variables $\hat A$ and $\hat m_g^2$ are related to each other for a given coupling $\lambda$ as the amount of screening is related to the occupancy $\hat A$. For the distribution given by \Eq{eq:kzdistribution},
\begin{align}\label{eq:mgAhat}
    \hat m_g^2  \approx \sqrt{\frac{3}{32\pi} } \lambda \hat A,
\end{align}
as is evident from \Eq{eq:mg}. 

\begin{figure*}
    \centering
\includegraphics[width=0.49\linewidth]{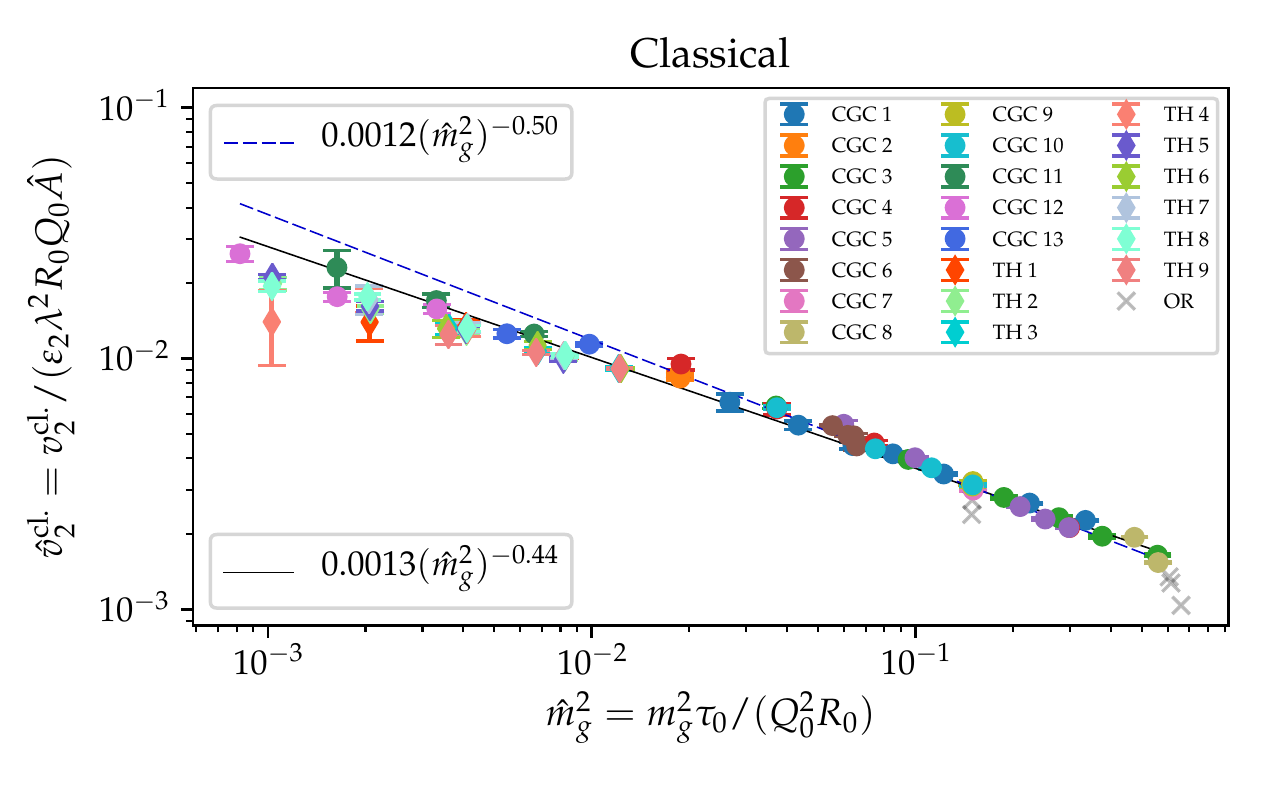}%
\includegraphics[width=0.49\linewidth]{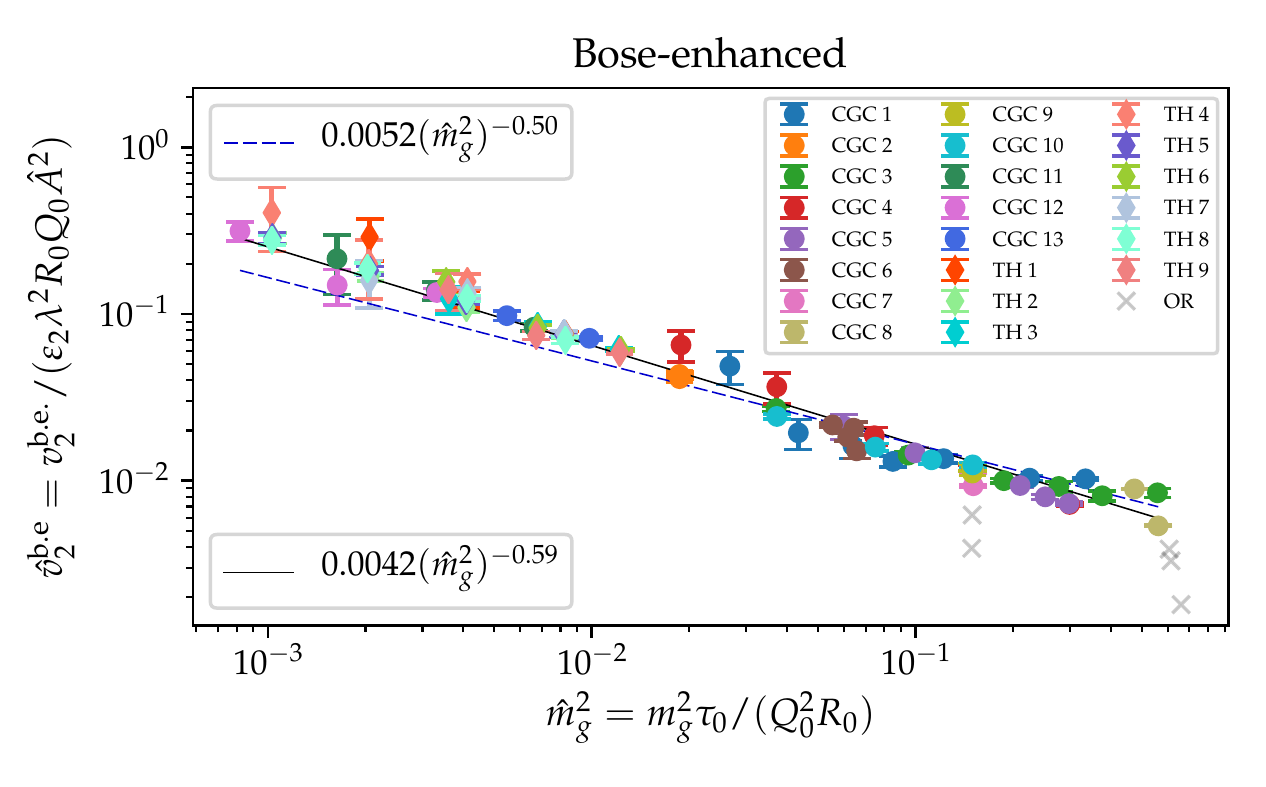}
    \caption{a) The scaling curve of the energy weighted elliptic flow for the classical part $\tilde{v}_2^\text{cl.}$ of $v_2$, \Eq{eq:v2scaling:second}. Labels containing CGC correspond to simulations with the CGC inspired initial condition defined in \Eq{eq:kzdistribution}. Labels containing TH are correspondingly for simulations with a deformed thermal initial distribution, \Eq{eq:thermaldistribution}. The black solid line is a power-law fit, and the dashed blue line is a fit with fixed power $-0.5$ for comparison. Points marked OR indicates simulations with parameters outside the range  of scaling regime and are not included in the fit. b) Analogous plot for the Bose-enhanced part of the energy weighted elliptic flow $\tilde{v}_2^\text{b.e.}$.}
    \label{fig:bose}
\end{figure*}

Our final scaling formula for the linear flow response in single-hit EKT,
\begin{align}
    &v_n/\varepsilon_n =  \hat R \left[ \hat v_n^\text{cl.}(\hat m_g^2) + \hat A \, \hat v_n^\text{b.e.}(\hat m_g^2)\right],
 \label{eq:v2scaling:second}
\end{align}
depends on two unknown functions $\hat v_n^\text{cl.}(\hat m_g^2)$ and $ \hat v_n^\text{b.e.}(\hat m_g^2)$. In addition, we identify the overall factor $\hat R = \lambda^2 R_0 Q_0 \hat A$ as a parametric estimate of the system size in units of the (transport, large-angle) mean free path at the origin of the system, $r=0$, at the time $\tau = R_0$ when the flow is built up,
\begin{align}\label{eq:Rhat}
\frac{ R_0}{l_{\rm mfp}} & \sim R_0 \, \sigma n  \sim R_0 \left[ \frac{\lambda^2}{Q_0^2}  \right]  \left[ A \frac{Q_0^3}{\xi R_0/\tau_0 } \right] \equiv \hat R,
\end{align}
where $\xi R_0/\tau_0$ is the anisoptropy of the distribution at time $\tau = R_0$.
Alternatively, we can understand \Eq{eq:v2scaling:second} in terms of initial gluon multiplicity. For the CGC-type initial conditions  (see \Eq{eq:CGCscaling}) and constant $\lambda$, we have that $\hat m_g^2 \propto \hat {A} \propto (dN_g/d\eta)^{-1/2}$, while $\hat{R}$ is approximately independent of multiplicity.

The classical  $\hat v_n^\text{cl.}(\hat m_g^2)$ and bose-enhanced $ \hat v_n^\text{b.e.}(\hat m_g^2)$ scaling functions can be extracted from numerical EKT simulations.
For the elliptic flow response $n=2$,  we performed 83 independent EKT simulations and systematically scanned the  parameter space of 
$(\tau_0/R_0,A,Q_0 R_0,\xi,\lambda)$ 
around our default choice 
$(0.1,4,18,4,10)$. We considered both CGC-like \Eq{eq:kzdistribution} and deformed thermal \Eq{eq:thermaldistribution} distribution functions (in the latter case, we used \Eq{eq:mgAhat} to define the $\hat{A}$ scaling variable in \Eq{eq:v2scaling:second}).  We tabulated different parameter choices in  Tables~\ref{tab:scalingCGC} and \ref{tab:scalingTH} together with the resulting classical and Bose-enhanced elliptic flow (see \app{sec:tables}).

In \Fig{fig:bose} we display the
classical  $\hat v_n^\text{cl.}(\hat m_g^2)$ and the  bose-enhanced $ \hat v_n^\text{b.e.}(\hat m_g^2)$ scaling functions as 
a function of $\hat{m}_g^2$. 
We observe that in both left and right panels different simulations collapse to a single universal curve within the statistical uncertainties. This demonstrates that the scaling form of \Eq{eq:v2scaling:second} derived in $\tau_0\ll R_0$ and $\xi\ll1$ is born out of realistic EKT simulations. This is a highly non-trivial validation of the scaling formula.
In order to obtain a pocket formula for the EKT response, we perform a linear fit in the loglog plot, which reasonably well describes the scaling functions. We note that for parameter choices far from this scaling regime (see Table~\ref{tab:scalingOR}), the data points do not fall on the same universal curve. Such points outside of range of the scaling regime (labelled as OR for ``Outside of Range") are shown in gray in \Fig{fig:bose}, but are not included in the fit.
We note that although the scaled classical part in \Fig{fig:bose} is numerically smaller than the Bose-enhanced term, the latter is normally suppressed by $\hat{A}\ll1$ in \Eq{eq:v2scaling:second}. For example, for $\lambda=10$, one needs $\hat{m}_g^2>0.6$ to  get $\hat{A}>1$.

\subsection{Comparison to Isotropization Time Approximation}
\label{sec:EKTvsRTA}

In the Relaxation Time Approximation (RTA) the collision kernel is taken to be~\cite{PhysRev.94.511}
\begin{equation}\label{eq:CITA}
   - C[f] = -\gamma e_{\rm rest}^{1/4} \frac{\bar E_\p}{p}(f(\tau,\xp;\p)-f_\text{iso}(\bar E_\p; e_{\rm rest}))
\end{equation}
where $\gamma$ is the single parameter controlling the rate of relaxation, $e_{\rm rest}$ is the (rest-frame) energy density in the frame moving with velocity $u^\mu$, which solves the Landau matching condition: $T^{\mu\nu}u_\nu =- e_{\rm rest} u^\mu$, and $\bar E_\p= -p^\mu u_\mu$ is the particle energy in that frame.
The term $f_\text{iso}(\bar E_\p; \varepsilon)$ is an isotropic thermal distribution towards which the system is relaxing, e.g., Boltzmann or Bose-Einstein distributions. The evolution of the energy flow in RTA does not depend on the $p$-dependence of $f_\text{iso}$, but only on the energy density, rest-frame and the assumption that $f_\text{iso}$ is isotropic. Therefore it is sufficient to study the momentum integrated Boltzmann equation and only keep track of the \emph{isotropization} in the Isotropization Time Approximation (ITA)~\cite{Kurkela:2018ygx}.
In this section we  bring our EKT results in contact with previous studies of 2D kinetic theory using ITA kinetic theory~\cite{Kurkela:2018ygx,Kurkela:2018qeb,Kurkela:2019kip,Kurkela:2019set,Kurkela:2020wwb}.

In ref.~\cite{Kurkela:2018ygx} it was noted that the ITA flow response is uniquely determined by the opacity parameter $\hat{\gamma}\equiv\gamma (e_\text{rest}\tau)_0^{1/4} R_0^{1/4}$, which has the interpretation of the system size in units of the mean-free-path, c.f. $\hat{R}$ in \Eq{eq:Rhat}.
A fair comparison between EKT and RTA should then be done for the same mean-free-path length. However, it is not trivial to relate the normalization of $l_\text{mfp}$ in the two kinetic theories and for simplicity we will compare ITA and EKT at fixed $\hat{\gamma}$.
To make a connection with EKT, we replace $\gamma$ with the specific shear viscosity $\eta/s$ using the equilibrium relation in ITA kinetic theory: $\gamma=\frac{T}{e_\text{rest}^{1/4}} \frac{1}{5\eta/s}$. For massless Bose particles with the initial distribution in \Eq{eq:kzdistribution} we can write ($e_\text{rest}=e_0$ initially)
\begin{equation}\label{eq:gammahat}
 \hat{\gamma} =\frac{1}{5\eta/s} \left(\frac{e_0}{\nu_g\pi^2/30}\frac{\tau_0}{R_0}\right )^{1/4}R_0 \approx 0.11 \frac{\hat{A}^{1/4}}{\eta/s} R_0Q_0.
 \end{equation}
In the linear regime, the ITA response for the integrated harmonic flow is given by the coefficients~\cite{Kurkela:2018ygx}
\begin{equation}\label{eq:ITAres}
    \frac{v_2^\text{ITA}}{\varepsilon_2\hat{\gamma}}=0.212, \quad
    \frac{v_3^\text{ITA}}{\varepsilon_3 \hat{\gamma}}=0.062. 
\end{equation}
We have implemented the RTA collision kernel, \Eq{eq:CITA}, and verified that we reproduce the results above obtained by the momentum-integrated Boltzmann equation in ITA.

\begin{figure}
    \centering
    \includegraphics[width=0.8\linewidth]{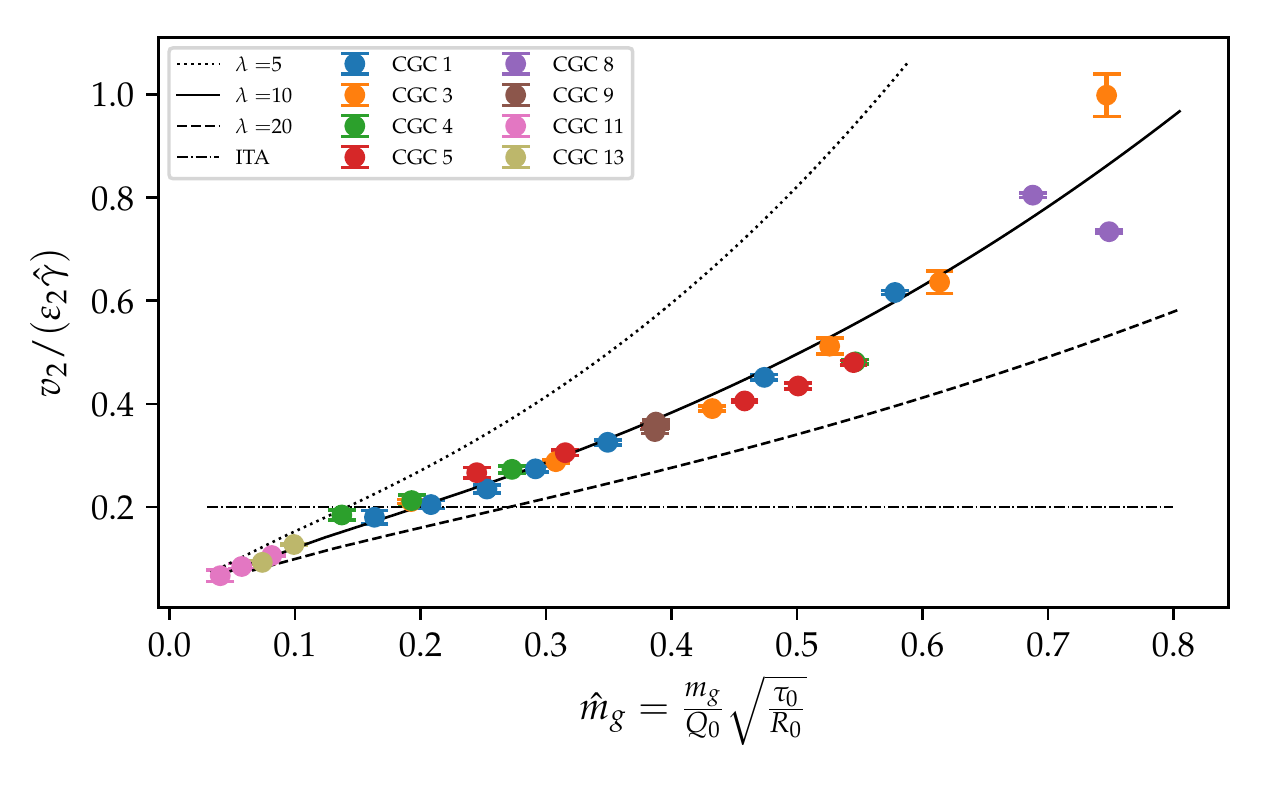}
    \caption{Comparison of $v_2/(\varepsilon_2\hat{\gamma})$ as a function of $\hat{m}_g$ for EKT and ITA, with fixed $\lambda=10$. The solid black lines are approximations of $v_2/(\varepsilon_2\hat{\gamma})$ obtained through a combination of the linear fits derived in Fig.~\ref{fig:bose}, for varying $\lambda$.}
    \label{fig:EKTvsITA}
\end{figure}
The collision kernel of RTA in \Eq{eq:CITA} is much simplier than the EKT collision kernel, \Eq{2to2}. In particular, the linear flow response dependence on the occupation is only $\hat A^{1/4}$, see \Eq{eq:gammahat}, in contrast to \Eq{eq:v2scaling:second}, which depends non-trivially on $\hat m_g^2$.
In \Fig{fig:EKTvsITA} we show the elliptic flow response $v_2/(\varepsilon_2 \hat{\gamma})$ in EKT as a function of $\hat{m}_g$.
The points correspond to EKT simulations for $\lambda=10$. The black lines are the results obtained by summing the power-law fits
in \Fig{fig:bose} for different values of $\lambda=5,10,20$ with $\eta/s\approx 2,0.6,0.2$.
For reference, we display the ITA value by a horizontal dot-dashed line.
The ratio with the opacity $\hat{\gamma}$ partially, but not completely, cancels the strong dependence on the coupling constant $\lambda$ in \Eq{eq:v2scaling:second}.
The EKT result for  $v_2/(\varepsilon_2 \hat{\gamma})$ grows approximately linearly with $\hat m_g$ and for $\hat m_g\approx 0.15$ it is roughly equal to the corresponding ITA values. At small values of $\lambda$ or small $\hat{A}$ values the EKT can be less efficient in generating elliptic flow than ITA at the same $\hat{\gamma}$ value. In the next section we will see that for initial conditions found in small collision systems, the EKT response is similar to that in ITA.

\begin{figure}
    \centering
    \includegraphics[width=1\columnwidth]{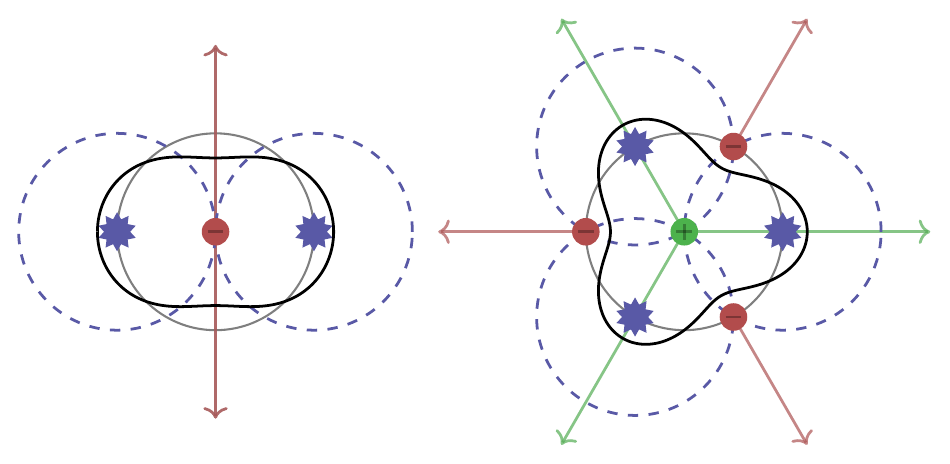}
    \caption{
    A simple cartoon illustrating the pertinent features of elliptic and triangular flow generation in the single-hit approximation. The initial anisotropy can be crudely modelled by 2 or 3 hot spots (blue stars), and the free streaming particles can be followed; the dashed circles show the location of the particles at some later time. For $v_2$, a majority of the flow is generated around the origin (red point), where the local isotropisation of the right and left movers leads to a global excess of up and down movers. As the momentum space anisotropy at the origin is aligned with the initial spatial anisotropy, the resulting momentum space anisotropy is anti-aligned with the spatial anisotropy.  For $v_3$, in addition to the origin, particles scatter also at $r > R_0$ at locations marked by the red dots. The local anisotropy in the origin is anti-aligned with the initial spatial anisotropy leading this time to  an aligned momentum space anisotropy. At the red dots, the rest frame is moving away from the collision zone. The collisions at red dots lead to an excess of particles moving along with the rest frame leading to an anti-aligned contribution. For $v_3$ the two regions (green dot, red dots) approximately cancel each other, leading to a small total contribution. 
    }
    \label{fig:flowgeometry}
\end{figure}

We close this section with a comment on the triangular flow.
All of the discussion for linear elliptic flow response also generalizes to arbitrary $v_n$ harmonic and, in particular, the triangular flow $n=3$. However, numerically, it is more difficult to study the scaling properties of $v_3$ response due to large numerical cancellations in the integral of $D_n(\tau,r)$ in \Eq{eq:vn}. 
A heuristic argument for the difference between the elliptic and triangular flow generation is given in the caption of  \Fig{fig:flowgeometry} for a toy example with $n=2$ or $n=3$ point sources.

\begin{figure*}
    \centering
\includegraphics[width=0.45\linewidth]{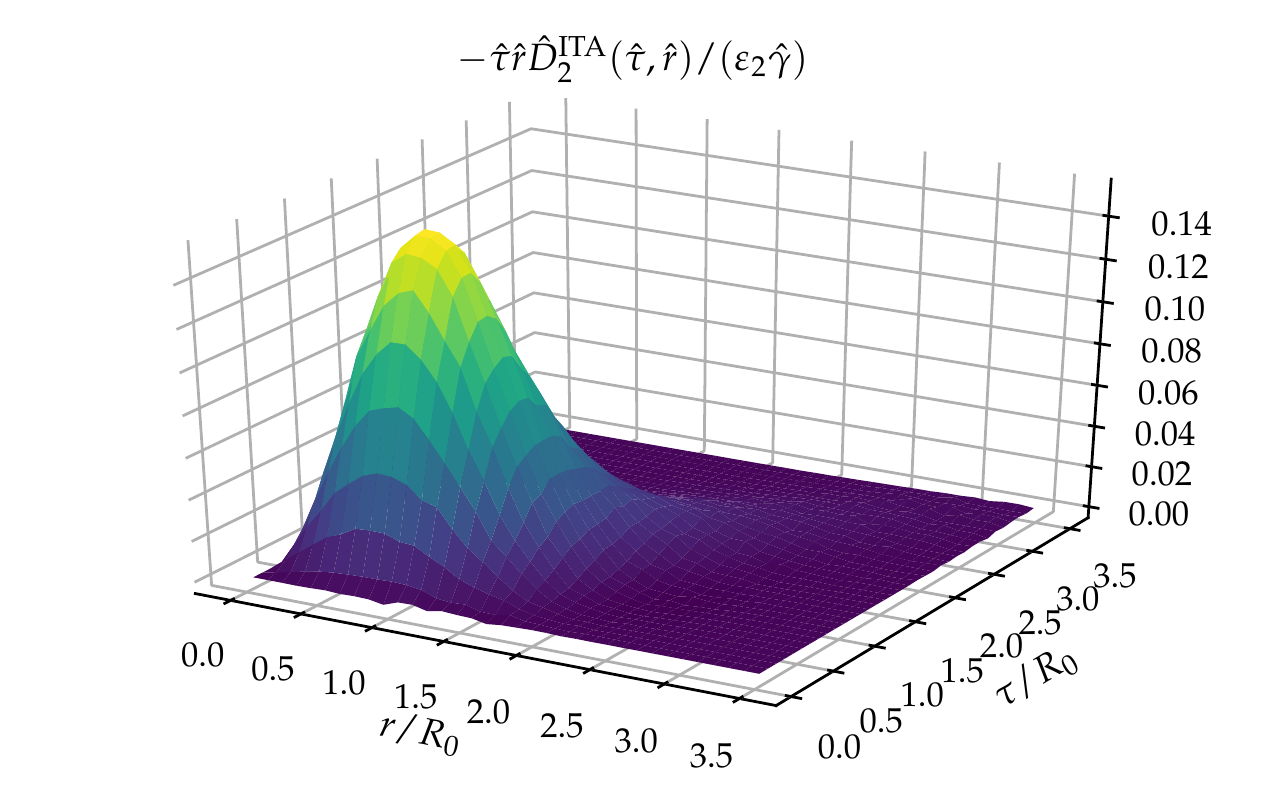}%
\includegraphics[width=0.45\linewidth]{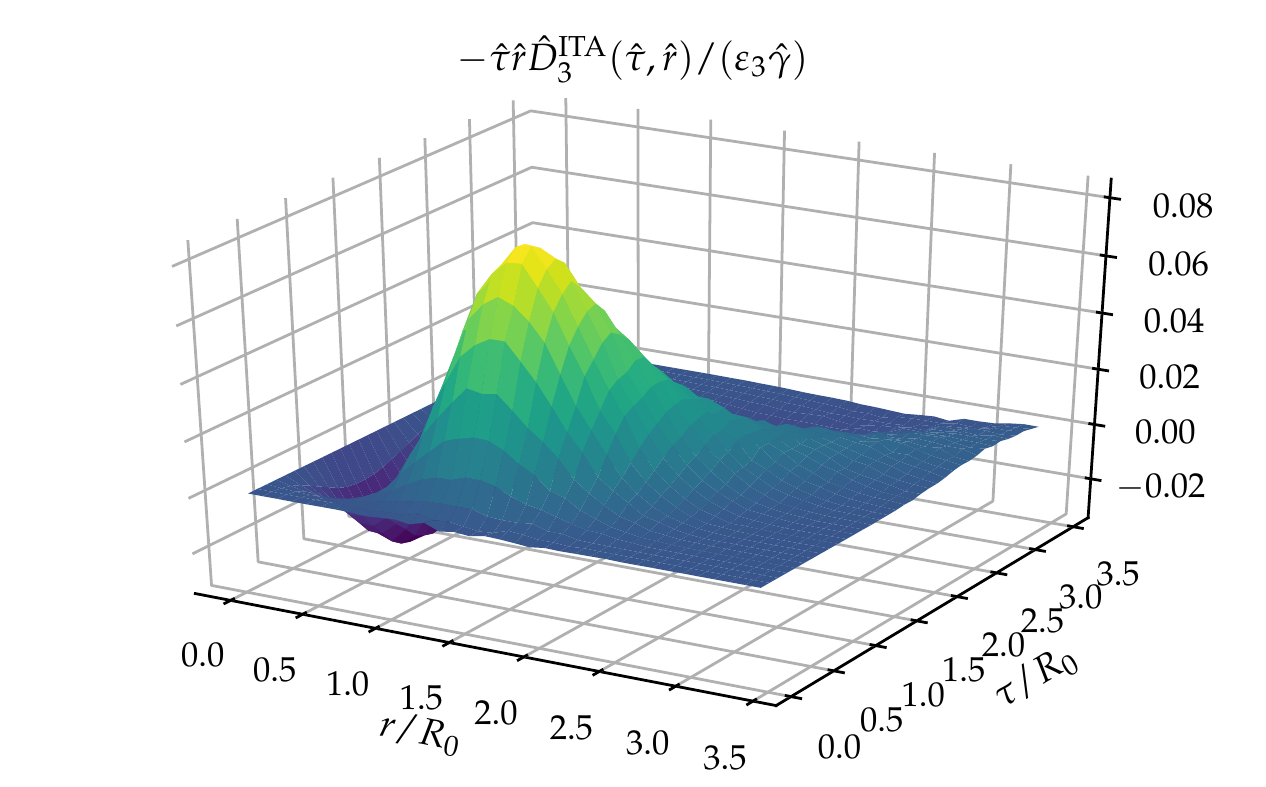}
\includegraphics[width=0.45\linewidth]{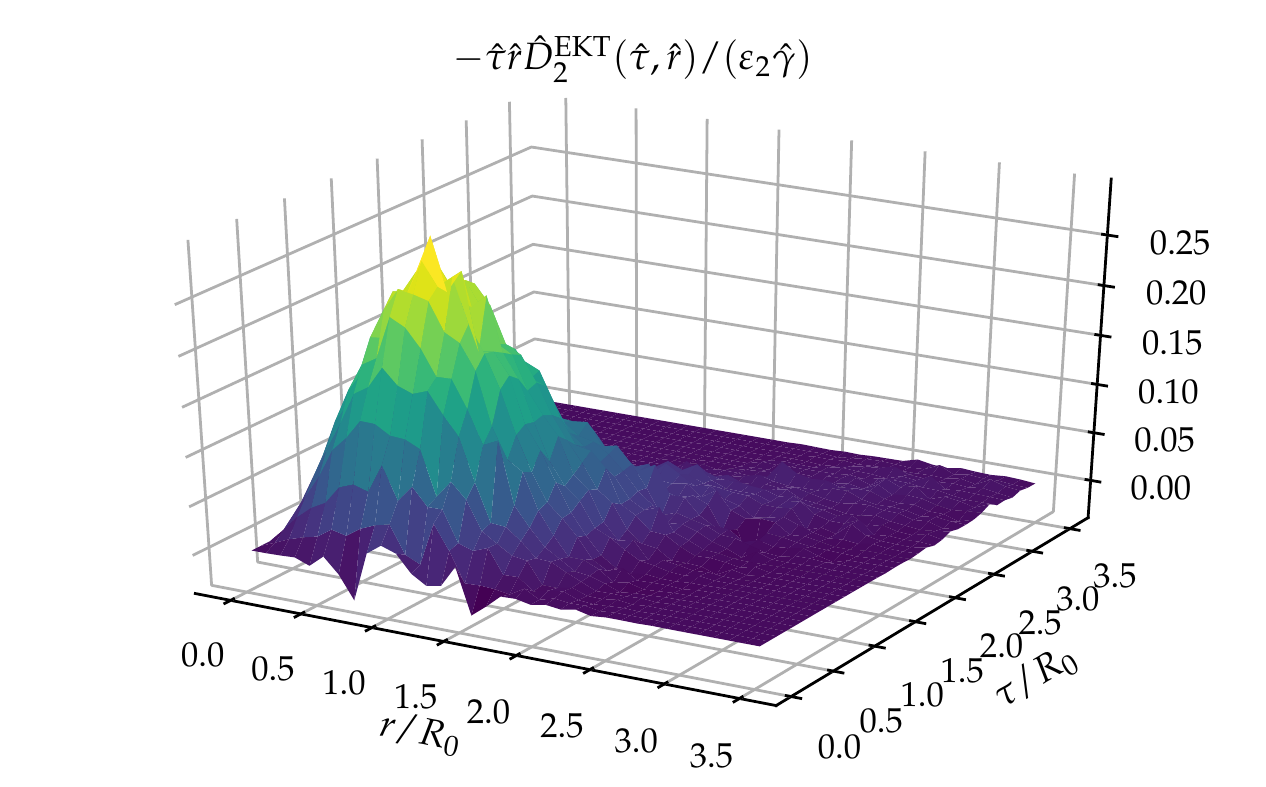}%
\includegraphics[width=0.45\linewidth]{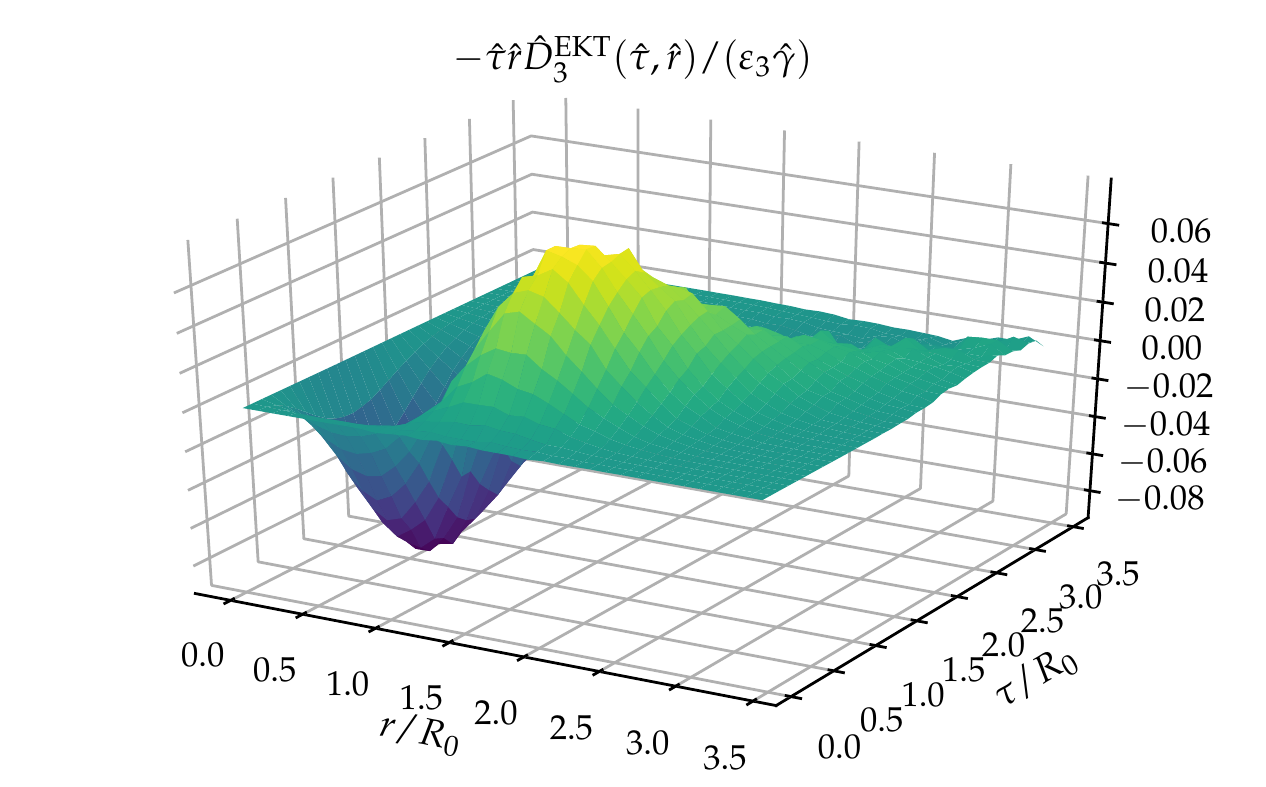}
    \caption{(Top row) Examples of differential distribution of energy weighted elliptic (left) and triangular (right) flow response $v_n/(\varepsilon_n \hat{\gamma})$ in the $\hat \tau\text{-}\hat r$ plane for ITA, see \Eq{eq:hatDn2}. (bottom row) Examples of energy weighted elliptic and triangular flow distributions in EKT simulations.}
    \label{fig:v3D}
\end{figure*}

\begin{figure*}
    \centering
    \includegraphics[width=0.49\linewidth]{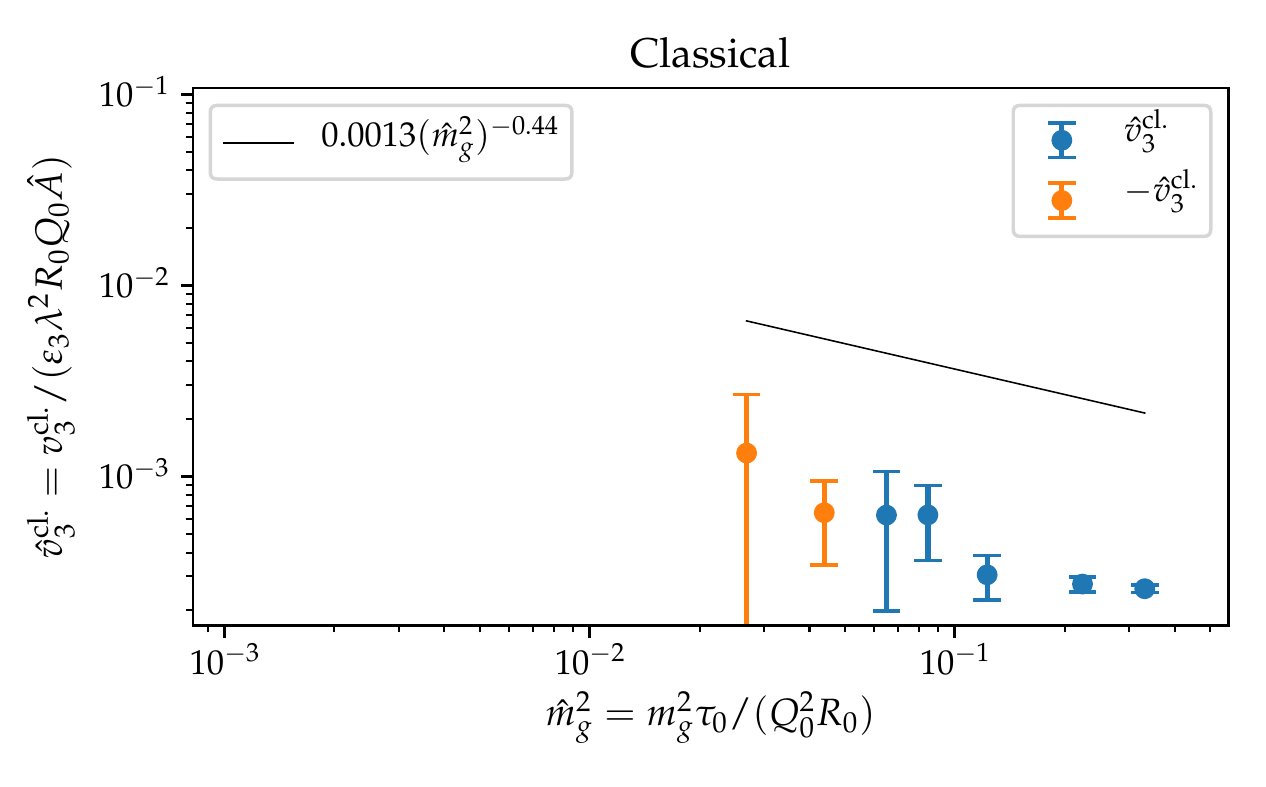}
    \includegraphics[width=0.49\linewidth]{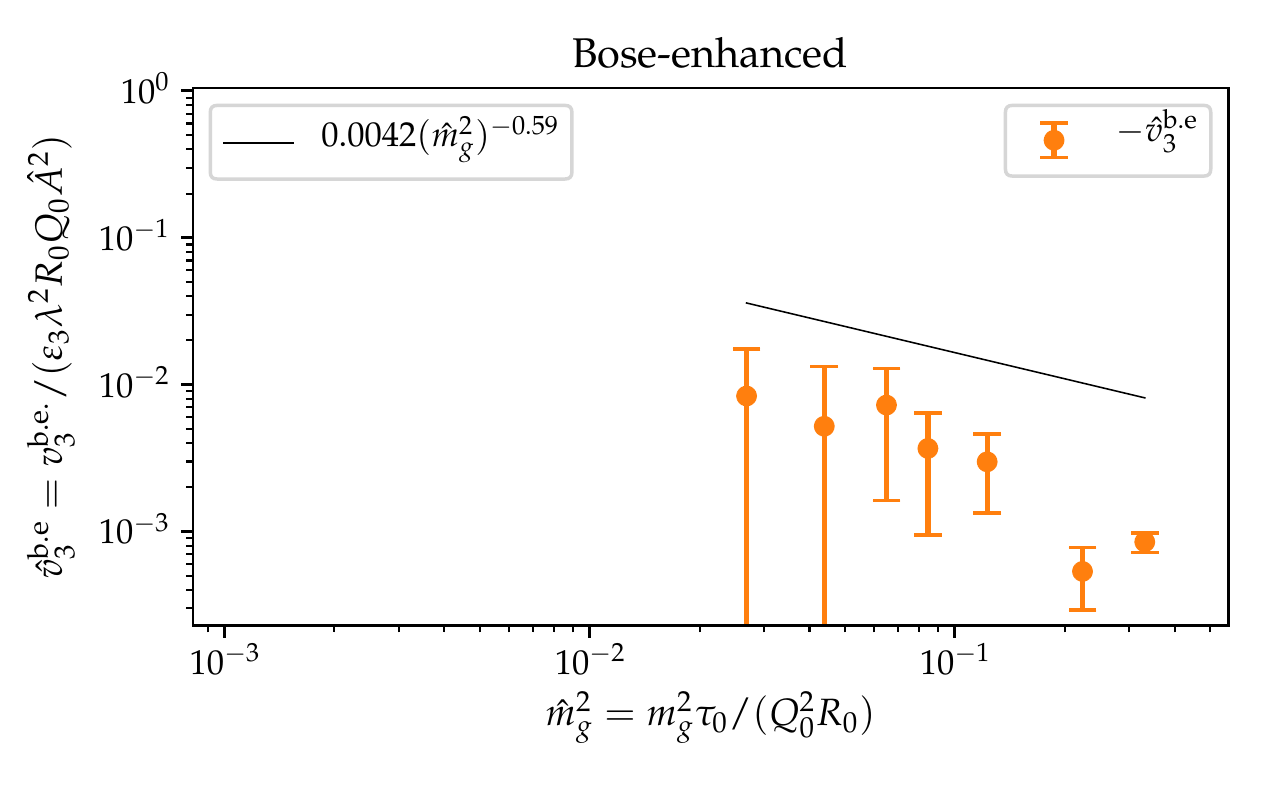}
    
    \caption{a) The energy weighted triangular flow for the classical part $\hat{v}_3^\text{cl.}$ of $v_3$, \Eq{eq:v2scaling:second} using CGC1 initial conditions defined by \Eq{eq:kzdistribution} and Table~\ref{tab:scalingCGC}. The power-law fit to elliptic flow results is shown for comparison as a black line. b) Analogous plot for the Bose-enhanced part of the energy weighted triangular flow $\hat{v}_3^\text{b.e.}$.}
    \label{fig:bosev3}
\end{figure*}

In \Fig{fig:v3D} we display the results for the
differential $v_n/(\varepsilon_n \hat{\gamma})$  distribution, i.e. $-\hat{r}\hat{\tau}\hat D_n(\hat \tau,\hat r)/(\varepsilon_n \hat{\gamma})$, for $n=2,3$ harmonics in ITA (top row) and EKT (bottom row). The integral of these distributions are equal to $v_2/(\varepsilon_2 \hat{\gamma})$ and $v_3/(\varepsilon_3 \hat{\gamma})$ correspondingly.
Indeed, we observe that for $v_2$ the distribution peaks at $r\approx\tau\approx R_0$ with positive contributions for all $r$ and $\tau$ values both for ITA and EKT.  In contrast, the negative triangular flow response $v_3/\varepsilon_3$ is generated at small radii $r\lesssim R_0$ and positive only for $r \gtrsim R_0$. Therefore there are significant cancellations between the two regions and the net response is small. In ITA the positive component is dominant and we obtained a net positive  $v_3/(\varepsilon_3 \hat{\gamma})$. However, in EKT, the contributions from large radii are smaller and nearly perfectly cancel the negative component at small radii. The end result is that $v_{3}/(\varepsilon_3\hat{\gamma})$ has a very small negative value. 
Finally, for completeness, in \Fig{fig:bosev3} we show triangular flow results for a particular set of CGC-type initial conditions. We see that both the classical and Bose-enhanced parts are an order of magnitude smaller than the corresponding terms for the elliptic flow.

\subsection{Energy weighted elliptic flow in small systems from single-hit EKT}

In this section we apply the scaling laws for EKT flow response extracted in \Sect{sec:conformalscaling} to realistic situations that take place in ultra-relativistic collisions of light nuclei, proton-nucleus and proton-proton collisions. In order to do so, we will use successful initial state, equilibration and hydrodynamic models to determine realistic initial conditions in small collision systems. We will then ask, what would be the expected $v_2$ signal if the system with the same initial conditions was to evolve in the single-hit EKT approximation. We emphasize that in this work we do not attempt to provide a complete description of signals of collectivity observed in small systems, as it clearly requires a detailed study of multiple observables. Rather, this is a proof-of-principle study of how efficient single-hit EKT is in generating elliptic flow signals.

The dynamical response of the EKT for a set of scaling variables is fully described by the scaling formula \Eq{eq:v2scaling:second}.
The combined power-fit functions results in the following pocket formula for the elliptic flow
\begin{align}
    v_2\approx \varepsilon_2 \hat{R} \left[ 1.3(\hat m_g^2)^{-0.44} + \hat A 4.2(\hat m_g^2)^{-0.59}\right]\cdot 10^{-3},
\end{align}
and the flow harmonic is given by the initial eccentricity $\varepsilon_2$, dimensionless system size $\hat R$, and dimensionless $\hat{A}$ and $\hat{m}_g^2$. These scaling parameters can be related to the physical dimensionful parameters ($R_0, (e\tau)_0$, and $Q_0$) via equations in \Sect{subsection:initialconditions}.

\begin{figure*}
    \centering
    \includegraphics[width=\linewidth]{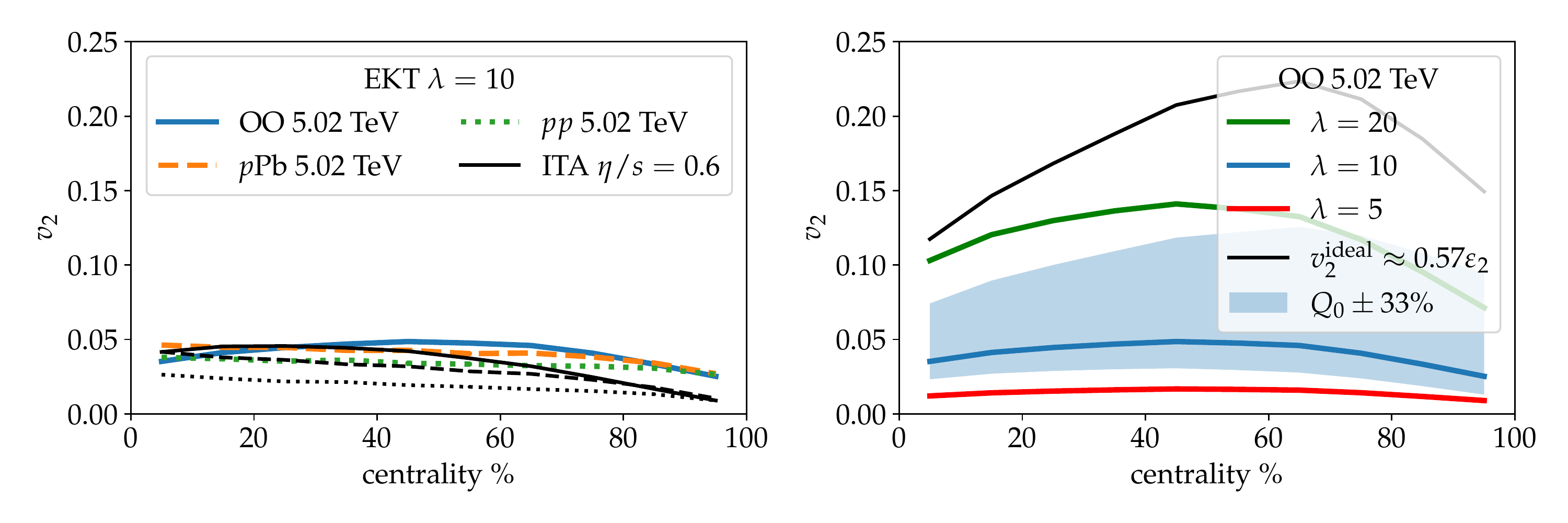}
    \caption{Left: Single hit EKT prediction for energy weighted elliptic flow in different small collision systems. Right: EKT response for two different values of the coupling constant (lines) and the sensitivity to the choice of $Q_0$ for $\lambda = 10$ (band).}
    \label{fig:v2pred}
\end{figure*}

We determine $R_0$ and $(e\tau)_0$ for different collision systems from the tabulated values of eccentricity $\varepsilon_2$, RMS radius $\sqrt{\left<r^2\right>}$ and entropy density $dS/dy/(\pi \left<r^2\right>)$ (in arbitrary units)  from ref.~\cite{Huss:2020whe}, that were generated using the TRENTo initial state model~\cite{Moreland:2014oya,Moreland:2018gsh}.
We summarize the procedure of obtaining initial conditions and tabulate the parameter values in \app{sec:initialtables}.

The knowledge of $R_0$ and $(e\tau)_0$ values is not, however, sufficient to uniquely determine $Q_0$ and $\hat{A}$. This is because $(e\tau)_0 \propto \hat{A} Q_0^4 R_0$ depends only on the combination $\hat{A} Q_0^4$ and to break this degeneracy we need to also provide an estimate for the typical momentum scale $Q_0$. In the saturation framework~\cite{Mueller:1999fp,Kovchegov:2000hz}, the initial gluon multiplicity per unity rapidity 
and area is $\propto Q^2_0$ (see \Eq{eq:CGCNg}). The mean transverse momentum of such gluons is of the order of the saturation scale $Q_0$. Therefore the initial gluon energy density in CGC-type initial conditions is
\begin{equation}
    (e\tau)_0\propto \left<p_T\right> \frac{dN_g}{d^2\x_\perp dy}\propto Q^3_0.\label{eq:etau0CGC}
\end{equation}
We fix the proportionality coefficient in \Eq{eq:etau0CGC} by choosing (somewhat arbitrarily) $Q_0$ to be equal to $3\,\text{GeV}$ at 0-10\% most central PbPb collisions at $\sqrt{s_\text{NN}}=5.02\,\text{TeV}$.
We will vary this value between $2\text{--}4$\,GeV to quantify the uncertainty arising from this choice.
Providing $Q_0$ allows us to determine what the scaling variables would be in a central PbPb collision. We use the estimated PbPb values $R_0=3.49\,\text{fm}$ and $(\tau e)_0=3.58\,\text{GeV}^3$ to find that this energy density corresponds to the initial distribution \Eq{eq:kzdistribution} with $\hat{A}_0=A\tau_0/(\xi R_0)\approx 0.00482$. Lastly, in order to extrapolate to other centralities and other collisions systems we rescale $\hat{A}$ and $Q_0$ values for PbPb using \Eqs{eq:etau0} and  \eq{eq:etau0CGC},
\begin{align}
  \frac{Q_0}{3.0\,\text{GeV}}&=  \left(\frac{(e\tau)_0}{3.58\,\text{GeV}^{3}}\right)^{1/3},\\
   \frac{\hat{A}}{0.00482}&=  \left(\frac{(e\tau)_0}{3.58\,\text{GeV}^{3}}\right)^{-1/3}\left(\frac{R_0}{3.49\,\text{fm}}\right)^{-1}.
\end{align}
These relations yield decreasing $Q_0$, but increasing $\hat{A}$ in smaller collision systems.

In \Fig{fig:v2pred} we illustrate the single-hit EKT response for elliptic flow in $\sqrt{s_\text{NN}}=5.02\,\text{TeV}$ OO, $p$Pb and $pp$ collisions. These are small collision systems for which a single-scattering approximation might be more appropriate than an infinite rescattering limit found in an ideal hydrodynamic description.
In the left panel we show the centrality dependence of $v_2$ in different systems for $\lambda=10$. $Q_0$ decreases from $\approx 1.5\,\text{GeV}$ in central to $\approx 0.3\,\text{GeV}$ in peripheral collisions for all three systems (see Tables~\ref{tab:OO_input}, \ref{tab:pPb_input} and \ref{tab:pp_input}). Correspondingly $\hat{A}$ increases from $\approx 0.02$ to $\approx 0.2$.
The single-hit EKT response per eccentricity and system size, i.e. $v_2/(\varepsilon_2 R_0Q_0)$, is a function of $\hat{m}_g^2\propto \hat{A}$, with stronger response for larger $\hat{A}$ (see \Fig{fig:EKTvsITA}). Even though in more peripheral bins for OO the system size and $Q_0$ reduces, the increase in $\hat{A}$ and $\varepsilon_2$ results in  a weak centrality dependence of net $v_2$.
For comparison we also show the ITA response,  \Eq{eq:ITAres}, for the same $\eta/s=0.6$ value.
In the ITA case $v_2/(\varepsilon_2 \hat{\gamma})$ is constant and therefore the $v_2$ becomes small in peripheral bins. We find that for $\lambda=10$ and $Q_0=3\,\text{GeV}$ the EKT and ITA response are similar in magnitude in central OO collisions, but deviate in peripheral bins.

 In the right panel of \Fig{fig:v2pred} we study the sensitivity of the EKT response to the coupling constant $\lambda$ and scale $Q_0$ in OO collisions. Green, blue and red lines correspond to $\lambda=20,10,5$ respectively and we see a strong dependence on the coupling constant.
For $\lambda=10$, we also show the blue band obtained by varying $Q_0$ by $33\%$ (i.e. varying $Q_0$ between 2 GeV and 4 GeV in a central PbPb collision). 
It is clear that the $\lambda$ and $Q_0$ values in elliptic flow response are degenerate. 
In addition we show the centrality dependence of the initial geometry eccentricity scaled by the ideal (and conformal) hydrodynamic response~\cite{Kurkela:2020wwb}. This illustrates the diametrically opposite limit of infinite rescatterings. As flow then follows the eccentricity, we observe that it is larger in more peripheral bins and is generally larger than the single hit EKT response for given values of $\lambda$ and $Q_0$.

Experimentally measured charged particle number elliptic flow in  $p$Pb and $pp$ collisions is only weakly dependent on centrality and of typical size $v_2\{2\}\approx 0.06$~\cite{Acharya:2019vdf}.
We see in \Fig{fig:v2pred} that for our choice of input parameters, the EKT single-hit approximation reproduces the same order of magnitude of the elliptic energy flow. Clearly, one should not expect that our simplified model would accurately describe the experimental data. It only demonstrates that approaches based on QCD effective kinetic theory can efficiently generate sizable harmonic flow even in the limit of few rescatterings.

\section{Conclusions and Outlook}
In this work we presented the first study of system-size dependence of harmonic flow response in QCD effective kinetic theory. We used the single-hit approximation to calculate the linear response coefficient for energy weighted elliptic flow on top of a free-streaming background.
Despite the simplifying assumptions, our study addresses a number of new dynamical features of the system that were not accessible in previous toy models. Energy flow response in QCD EKT is generated via the elastic $2\leftrightarrow 2$ scatterings, which can be Bose-enhanced if the phase-space occupation density is large enough. In addition, the elastic scattering matrix element is regulated by the in-medium screening mass. This leads to a non-trivial scattering rate and flow response dependences on the initial conditions. We find the scaling laws that relate flow response with different initial conditions to each other and we  provide a simple pocket formula parametrization of the elliptic flow response in EKT.

We apply single-hit EKT response to estimate the centrality dependence of energy weighted elliptic flow in $\sqrt{s_\text{NN}}=5.02\,\text{TeV}$ OO, $p$Pb and $pp$ collisions. Although there are significant systematic uncertainties and simplifications involved, the resulting energy weighted elliptic flow for a realistic choice of parameters was found to be in order of magnitude agreement with the experimentally measured charged particle number elliptic flow in  $p$Pb and $pp$ collisions.

For the initial conditions studied in this paper both the EKT and ITA responses are rather similar for the energy weighted elliptic flow, but we found a much smaller energy weighted triangular flow in the EKT simulations than in those using ITA.
This indicates that further study of kinetic theories with more complicated collision kernels like EKT can lead to specific collective flow signatures that may allow to distinguish between different microscopic interaction mechanisms. One of the clear advantages of EKT over ITA, is that the momentum dependence of flow harmonics can be studied. However, to correctly describe the $p_T$-resolved $v_n$ one will need to take into account the  $p_T$ profile of initial conditions, the collinear processes and hadronization. 
Finally, going beyond the linearized single-hit approximation employed in this work will be important for connecting the small and large systems and studying the hydrodynamization as a function of the system size.

 {\bf Acknowledgements.}

We thank  Wilke van der Schee, Sören Schlichting,  Urs Wiedemann and Bin Wu for useful discussions.
RT work is funded by the European Research Council (ERC) under the European Union's Horizon 2020 research and innovation programme (grant agreement No 803183, collectiveQCD). RT thanks CERN Theoretical Physics Department for their hospitality during a short term visit.

\bibliographystyle{JHEP}
\bibliography{master}

\appendix

\section{Linearizing $D_n$}
\label{sec:Dn}
In this section we derive the explicit expression
for the time- and radius-resolved flow distribution  $D_n = -\frac{dv_n}{\tau d\tau r dr}$
given    in \Eq{eq:vnDn}.
    The linearized elastic collision kernel $\delta C_{2\leftrightarrow 2}$ contains two types of terms. The first of them is due to the linearization of the phase-space distributions in the loss and gain factors in \Eq{2to2}. Using the free-streaming solution, we can write for time $\tau\geq\tau_0$
    \begin{align}
      &  f^{(0)}(\tau,\xp; \p) = \bar f_\p + \delta f_\p\nonumber \\
        & =\bar{f}(\tau,|\txp|; \p)\left (1 + \epsilon \frac{|\txp|^n}{R_0^n}\cos(n \phitx)\right).
    \end{align}
    Here the $|\tilde \x_\perp|$ and $\phi_{\tilde \x}$ is the radius and angle of the co-moving coordinate, \Eq{eq:xtilde}.

    The second term in $\delta C_{2\leftrightarrow 2}$ arises due to linear variation of the gluon screening mass in the regulated $t_\text{reg}$ and $u_\text{reg}$, see \Eq{eq:treg}, in the matrix element \Eq{eq:matrixelement}.
    Splitting the screening mass in the background and perturbation $m^2_g = \bar{m}^2_g + \delta m^2_g$ and using \Eq{eq:m2} we can write that
    \begin{align}
        \delta m^2_g & = 2 \lambda \int \frac{d^3 p}{(2\pi)^3|\p|}\delta f_\p\nonumber \\
        & = \epsilon \cos(n\phi_\x)2\lambda \int \frac{d^3 p}{(2\pi)^3|\p|} \frac{|\tilde \x_\perp|^n}{R^n_0} \bar{f} \cos(n \phiDtx)\nonumber \\
        & \equiv \epsilon \cos(n\phi_\x) \delta m'^2_g(\tau_0,|\txp|).
    \end{align}
    In the second line we expanded the $\cos (n \phitx)=\cos(n\phix)\cos(n\phiDtx)-\sin(n\phix)\sin(\phiDtx)$, where $\phiDtx = \phitx-\phix$. We dropped the sin terms, because $|\tilde \x_\perp|$ is an even function in  the relative momentum angle $\phi_\p-\phi_\x$, while $\phiDtx$ is odd. Finally in the last line we defined $\delta m'^2_g$ and explicitly factored out the angular dependence on $\phix$. Then we can linearize the scattering matrix as
    \begin{align}
        \label{eq:M2_perturbation}
       & |\mathcal{M}(m^2_g)|^2  = |\mathcal{M}(\bar{m}_g^2)|^2+\delta|\mathcal{M}(\bar{m}_g^2,\delta m^2_g)|^2 \nonumber\\
        & = |\mathcal{M}(\bar{m}_g^2)|^2 + \epsilon \cos(n\phix)\delta|\mathcal{M}(\bar{m}_g^2,\delta m'^2_g)|^2.
    \end{align}
    With the definitions given above the first step of the linearization of $D_n$  is now straightforward  ($\int_\p\equiv \int \frac{d^3\p}{2p(2\pi)^3}$)
    \begin{align}
        \label{eq:Dn_first}
        D_n & = \left( \left.\frac{dE_\perp}{2\pi d\eta}\right|_{\bar{f}}\right)^{-1} \frac{1}{2\nu_g}\int \frac{d\phix}{2\pi} \int_{\p\k\pp\kp} p_\perp  \cos (n\phip) \nonumber\\
        & \times |\mathcal{M}(\bar{m}^2_g)|^2 (2\pi)^4 \delta^{(4)}(P+K-P'-K') 
        \nonumber \\
        &\times   \Bigg\{ \bar{f}_\p \bar{f}_\k[1+ \bar{f}_{\p'}][1+ \bar{f}_{\k'}]\Big\{\frac{\delta |\mathcal{M}(\bar{m}_g^2, \delta m^2_g)|^2}{|\mathcal{M}(\bar{m}^2_g)|^2}\nonumber \\
        & + \frac{\delta f_\p}{ \bar{f}_\p}+\frac{\delta f_\k}{ \bar{f}_\k}+\frac{\delta f_{\p'}}{ 1+ \bar{f}_{\p'}}+\frac{\delta f_{\k'}}{1+ \bar{f}_{\k'}} \Big\} \nonumber \\
        &-\bar{f}_{\p'}\bar{f}_{\k'}[1+ \bar{f}_{\p}][1+ 
        \bar{f}_{\k}] \Big\{\frac{\delta |\mathcal{M}(\bar{m}_g^2,\delta m^2_g)|^2}{|\mathcal{M}(\bar{m}^2_g)|^2} \nonumber \\
        & + \frac{\delta f_\p}{ 1+ \bar{f}_\p}+\frac{\delta f_\k}{1+ \bar{f}_\k}+\frac{\delta f_{\p'}}{ \bar{f}_{\p'}}+\frac{\delta f_{\k'}}{ \bar{f}_{\k'}} \Big\}\Bigg\}.
    \end{align}
    
    We now turn our attention to the angular dependence of the equation above. Both terms of the distribution, $\bar{f}$ and $\delta f$, have an angular dependence through the magnitude of the co-moving coordinate $|\txp|$. By squaring \Eq{eq:xtilde} we can write
    \begin{equation}
        |\txp|^2 = r^2+L^2-2 rL \cos(\phix-\phip) 
    \end{equation}
    with
    \begin{equation}
        L \equiv \tau \sqrt{1+\frac{p_z^2}{p_\perp^2}}- \tau_0\sqrt{1+\frac{p_z^2\tau^2}{p_\perp^2\tau_0^2}} .
    \end{equation}
    Hence, $|\txp|$ depends only on the relative angle $\phip-\phix$. The only other angular dependence appearing in \Eq{eq:Dn_first} are the explicit terms $\cos (n\phi_{\tilde \x(\k)})\cos (n\phip)$.
    Using again $\phiDtx = \phitx-\phix$, which is a function of the relative angle $\phik-\phix$, and writing $\phip=\phip-\phix+\phix$ we factor $\cos (n\phitx)\cos (n\phip)$ in terms depending on the relative momentum angles and $\phix$ as follows  
        \begin{align}\label{eq:cosexp}
         &\cos\left( n\phitx \right)\cos(n\phip)  =\nonumber\\
       &=\cos\left( n\phiDtx\right) \cos(n(\phip-\phix))\cos^2\left( n\phix\right)\nonumber \\
        &+\sin\left( n\phiDtx\right) \sin(n(\phip-\phix))\sin^2\left( n\phix\right)\nonumber\\
      &-\cos\left( n\phiDtx\right) \sin(n(\phip-\phix)) \cos\left( n\phix\right)\sin\left( n\phix\right)\nonumber\\
      &-\sin\left( n\phiDtx\right) \cos(n(\phip-\phix))    \cos\left( n\phix\right)\sin\left( n\phix\right).
     \end{align}
Hence, every term in the integral in \Eq{eq:Dn_first} depends only on the relative momentum angle or explicitly on $\phix$. Therefore, we can shift all four integration momentum angles by $\phix$ and eliminate $\phix$ everywhere except for the explicit terms. After doing the $\phix$ integral the cross-terms in \Eq{eq:cosexp} vanish, while terms with $\cos^2\phix$ and $\sin^2\phix$ add up to $\frac{1}{2}\cos(\phiDtx-\phip)$. Since the integral is symmetric in $\p$, $\k$, $\pp$ and $\kp$, except for $p_\perp$ and $\phi_\p$ in  $\cos(\phiDtx-\phip)$, we will symmetrize over all four momentum variables and divide the integral by $4$. Introducing a shorthand notation 
    \begin{align}
        C_{\p} & = \frac{|\txp (\p)|^n}{R_0^n}\Big\{ p_\perp \cos(n(\phi_{\Delta\tilde \x(\phip)}-\phip)) \nonumber \\
        & + k_\perp \cos(n(\phi_{\Delta\tilde \x(\phip)}-\phik)) - p'_\perp \cos(n(\phi_{\Delta\tilde \x(\phip)}-\phipp)) \nonumber\\ 
        & - k'_\perp \cos(n(\phi_{\Delta\tilde \x(\phip)}-\phikp)) \Big\}
    \end{align}
  we arrive at our final expression
    \begin{align}
        \label{eq:DNlinearfull}
        &D_n(\tau,r)  = \epsilon\frac{1}{16\nu_g}\left( \left.\frac{dE_\perp}{2\pi d\eta}\right|_{\bar{f}}\right)^{-1}\int_{\p\k\pp\kp}|\mathcal{M}(\bar{m}_g^2)|^2
          (2\pi)^4 \delta^{(4)}(P+K-P'-K')  \nonumber \\
          & \times \Bigg[ \frac{\delta |\mathcal{M}(\bar{m}_g^2, \delta m'^2_g)|^2}{|\mathcal{M}(\bar{m}^2_g)|^2}
        \Big\{ \bar{f}_\p \bar{f}_\k[1+ \bar{f}_{\p'}][1+ \bar{f}_{\k'}]-\bar{f}_{\p'}\bar{f}_{\k'}[1+ \bar{f}_{\p}][1+ 
        \bar{f}_{\k}]\Big\} \nonumber \\ 
         & \times ( p_\perp  \cos (n\phip) + k_\perp  \cos (n\phik) - {k'}_\perp  \cos (n\phikp)  - p'_\perp  \cos (n\phipp)  ) \nonumber\\
        &+ \Bigg\{ \bar{f}_\p \bar{f}_\k[1+ \bar{f}_{\p'}][1+ \bar{f}_{\k'}]\Big(\frac{ \bar{f}_{\p'}}{ 1+ \bar{f}_{\p'}}C_{\p'}+\frac{\bar{f}_{\k'}}{1+ \bar{f}_{\k'}} C_{\k'} + C_\p+C_\k\Big) \nonumber\\&\qquad-\bar{f}_{\p'}\bar{f}_{\k'}[1+ \bar{f}_{\p}][1+ 
        \bar{f}_{\k}] \Big( C_{\p'}+C_{\k'}\frac{ \bar{f}_\p}{ 1+ \bar{f}_\p}C_\p+\frac{ \bar{f}_\k}{1+ \bar{f}_\k}C_\k\Big)\Bigg\}\Bigg].
    \end{align}
The multidimensional integral present in the linearized form of \Eq{eq:DNlinearfull} has been evaluated numerically using Monte Carlo with importance sampling for discrete values of $r$ and $\tau$. For the explicit phase-space parametrization of the integral see ref.~\cite{Keegan:2015avk}.

Mean values and stochastic errors presented in the figures and tables throughout this article have been calculated using the jackknife resampling method. Discretization errors in the $r-\tau$ plane with a $N_r\times N_\tau=20\times 20$ grid were checked to be of percent size and negligible in comparison to the statistical uncertainties.

In order to verify the results of both analytical predictions and numerical simulations, a number of crosschecks have been preformed. The validity of the derivation and implementation of the linerized expression of $D_n$, \Eq{eq:DNlinearfull}, has been checked against a nonlinear implementation, i.e. evaluating \Eq{eq:vnfull} directly. However, the nonlinear method has larger statistical uncertainties, which become worse for very small values of $\epsilon$. Therefore all reported results are obtained with the linearized equation \Eq{eq:DNlinearfull}. We have also implemented RTA kinetic theory \Eq{eq:CITA} in the same setup. As both  kinetic theories share the same free-streaming background evolution, the reproducability ITA results in \Eq{eq:ITAres} was used as additional check.

\section{Results of the parameter space scan}
\label{sec:tables}
To numerically test the scaling relations predicted in \Sect{sec:conformalscaling} we performed a systematic variation of all model parameters: starting time $\tau_0$, Gaussian width of the density profile $R_0$, the normalization of the distribution $A$, anisotropy parameter $\xi$ and the coupling constant $\lambda$. We tabulate the values of the screening mass $m_g^2$ at initial time at the origin ($r_0=0.01$), scaling variable $\hat m_g^2$ and the classical and Bose-enhanced contributions to the elliptic flow. For the linearized approach, we can completely scale the value of the eccentricity $\varepsilon_n$. The results for the CGC-like momentum distribution, \Eq{eq:kzdistribution}, are given in Table~\ref{tab:scalingCGC}. In addition, we performed simulations with deformed thermal initial conditions, \Eq{eq:thermaldistribution} and the results are given in Table~\ref{tab:scalingTH}.  Finally, for completeness in Table~\ref{tab:scalingOR} we record the results of simulations for which we do not expect scaling, because $\tau_0/R_0\ll1$ assumption is violated.
These results were used to produce Figs.~\ref{fig:bose} and \ref{fig:EKTvsITA}.

\setlength{\tabcolsep}{5pt}

	\begin{table*}[h!]
	\centering
	\def\arraystretch{0.63}
	\setlength{\tabcolsep}{0.25em}
	\begin{tabular}{ c|c|c|c|c|c|c|c|c|c|c }
		Label & $\tau_0$ & $R_0$ & $A$ & $Q_0$ & $\xi$ & $\lambda$  & $m_g^2(r_0,\tau_0)$ & $\hat{m}_g^2$ & $v_2^\text{cl.}/\varepsilon_2$ & $v_2^\text{b.e.}/\varepsilon_2$ \\ 
		\hline		\hline 
 		\multirow{7}{3.2em}{\text{CGC 1}} 
			 & 1 & 10 & 4 & 1.8 & 1.5 & 10 & 14.33 & 0.3342 & $7.88(3)\cdot10^{-1}$ & $6.91(9)\cdot10^{-1}$\\
			 & 1 & 10 & 4 & 1.8 & 2.5 & 10 & 9.632 & 0.2246 & $6.21(4)\cdot10^{-1}$ & $3.2(1)\cdot10^{-1}$\\
			 & 1 & 10 & 4 & 1.8 & 5 & 10 & 5.229 & 0.1219 & $4.40(5)\cdot10^{-1}$ & $1.22(7)\cdot10^{-1}$\\
			 & 1 & 10 & 4 & 1.8 & 7.5 & 10 & 3.647 & 0.08504 & $3.69(8)\cdot10^{-1}$ & $5.7(4)\cdot10^{-2}$\\
			 & 1 & 10 & 4 & 1.8 & 10 & 10 & 2.743 & 0.06397 & $3.00(9)\cdot10^{-1}$ & $4.0(6)\cdot10^{-2}$\\
			 & 1 & 10 & 4 & 1.8 & 15 & 10 & 1.861 & 0.0434 & $2.45(9)\cdot10^{-1}$ & $2.2(4)\cdot10^{-2}$\\
			 & 1 & 10 & 4 & 1.8 & 25 & 10 & 1.144 & 0.02667 & $1.9(1)\cdot10^{-1}$ & $2.1(5)\cdot10^{-2}$\\
		\hline 
 		\multirow{4}{3.2em}{\text{CGC 2}} 
			 & 1 & 10 & 10 & 1.8 & 4 & 0.5 & 0.8038 & 0.01874 & $3.26(6)$ & $3.6(1)$\\
			 & 1 & 10 & 5 & 1.8 & 4 & 1 & 0.7996 & 0.01865 & $1.66(3)$ & $8.6(4)\cdot10^{-1}$\\
			 & 1 & 10 & 1 & 1.8 & 4 & 5 & 0.7991 & 0.01863 & $3.34(7)\cdot10^{-1}$ & $3.6(2)\cdot10^{-2}$\\
			 & 1 & 10 & 0.5 & 1.8 & 4 & 10 & 0.8023 & 0.01871 & $1.71(3)\cdot10^{-1}$ & $9.1(5)\cdot10^{-3}$\\
		\hline 
 		\multirow{6}{3.2em}{\text{CGC 3}} 
			 & 1 & 10 & 1 & 1.8 & 4 & 10 & 1.59 & 0.03706 & $2.50(5)\cdot10^{-1}$ & $2.25(8)\cdot10^{-2}$\\
			 & 1 & 10 & 2.5 & 1.8 & 4 & 10 & 4.064 & 0.09476 & $3.91(4)\cdot10^{-1}$ & $7.7(3)\cdot10^{-2}$\\
			 & 1 & 10 & 5 & 1.8 & 4 & 10 & 8.023 & 0.1871 & $5.45(7)\cdot10^{-1}$ & $2.11(7)\cdot10^{-1}$\\
			 & 1 & 10 & 7.5 & 1.8 & 4 & 10 & 11.87 & 0.2767 & $6.7(1)\cdot10^{-1}$ & $4.3(3)\cdot10^{-1}$\\
			 & 1 & 10 & 10 & 1.8 & 4 & 10 & 16.15 & 0.3766 & $7.7(1)\cdot10^{-1}$ & $6.9(5)\cdot10^{-1}$\\
			 & 1 & 10 & 15 & 1.8 & 4 & 10 & 23.92 & 0.5577 & $9.56(9)\cdot10^{-1}$ & $1.6(1)$\\
		\hline 
 		\multirow{5}{3.2em}{\text{CGC 4}} 
			 & 0.25 & 20 & 4 & 1.8 & 4 & 10 & 6.466 & 0.01885 & $3.7(2)\cdot10^{-1}$ & $2.8(6)\cdot10^{-2}$\\
			 & 0.5 & 20 & 4 & 1.8 & 4 & 10 & 6.387 & 0.03723 & $4.9(3)\cdot10^{-1}$ & $6(1)\cdot10^{-2}$\\
			 & 1 & 20 & 4 & 1.8 & 4 & 10 & 6.388 & 0.07448 & $7.2(2)\cdot10^{-1}$ & $1.2(1)\cdot10^{-1}$\\
			 & 2 & 20 & 4 & 1.8 & 4 & 10 & 6.394 & 0.1491 & $9.79(1)\cdot10^{-1}$ & $3.2(1)\cdot10^{-1}$\\
			 & 4 & 20 & 4 & 1.8 & 4 & 10 & 6.404 & 0.2987 & $1.319(8)$ & $7.8(2)\cdot10^{-1}$\\
		\hline 
 		\multirow{6}{3.2em}{\text{CGC 5}} 
			 & 1 & 5 & 4 & 1.8 & 4 & 10 & 6.371 & 0.2971 & $3.28(3)\cdot10^{-1}$ & $1.94(4)\cdot10^{-1}$\\
			 & 1 & 6 & 4 & 1.8 & 4 & 10 & 6.458 & 0.251 & $3.60(3)\cdot10^{-1}$ & $1.82(6)\cdot10^{-1}$\\
			 & 1 & 7 & 4 & 1.8 & 4 & 10 & 6.305 & 0.21 & $3.94(2)\cdot10^{-1}$ & $1.74(3)\cdot10^{-1}$\\
			 & 1 & 10 & 4 & 1.8 & 4 & 10 & 6.419 & 0.1497 & $5.00(6)\cdot10^{-1}$ & $1.55(4)\cdot10^{-1}$\\
			 & 1 & 15 & 4 & 1.8 & 4 & 10 & 6.397 & 0.09945 & $6.25(8)\cdot10^{-1}$ & $1.3(1)\cdot10^{-1}$\\
			 & 1 & 25 & 4 & 1.8 & 4 & 10 & 6.425 & 0.05993 & $8.5(3)\cdot10^{-1}$ & $1.2(2)\cdot10^{-1}$\\
		\hline 
 		\multirow{4}{3.2em}{\text{CGC 6}} 
			 & 1 & 10 & 4 & 1.8 & 2.5 & 2.5 & 2.372 & 0.05531 & $1.25(1)$ & $6.4(2)\cdot10^{-1}$\\
			 & 1 & 10 & 4 & 1.8 & 5 & 5 & 2.645 & 0.06167 & $6.35(7)\cdot10^{-1}$ & $1.68(9)\cdot10^{-1}$\\
			 & 1 & 10 & 4 & 1.8 & 7.5 & 7.5 & 2.761 & 0.06438 & $4.40(9)\cdot10^{-1}$ & $9.2(8)\cdot10^{-2}$\\
			 & 1 & 10 & 4 & 1.8 & 10 & 10 & 2.817 & 0.06568 & $3.1(1)\cdot10^{-1}$ & $3.9(4)\cdot10^{-2}$\\
		\hline 
 		\multirow{2}{3.2em}{\text{CGC 7}} 
			 & 2 & 10 & 4 & 1.8 & 4 & 5 & 3.227 & 0.1505 & $9.38(6)\cdot10^{-1}$ & $5.10(7)\cdot10^{-1}$\\
			 & 1 & 10 & 4 & 1.8 & 4 & 10 & 6.454 & 0.1505 & $4.83(7)\cdot10^{-1}$ & $1.59(7)\cdot10^{-1}$\\
		\hline 
 		\multirow{2}{3.2em}{\text{CGC 8}} 
			 & 1.25 & 10 & 4 & 1.8 & 1.25 & 10 & 16.24 & 0.4733 & $9.56(4)\cdot10^{-1}$ & $1.21(1)$\\
			 & 2.5 & 10 & 4 & 1.8 & 2.5 & 10 & 9.618 & 0.5607 & $8.98(3)\cdot10^{-1}$ & $1.016(9)$\\
		\hline 
 		\multirow{4}{3.2em}{\text{CGC 9}} 
			 & 0.5 & 5 & 4 & 1.8 & 4 & 10 & 6.439 & 0.1502 & $2.46(4)\cdot10^{-1}$ & $8.2(4)\cdot10^{-2}$\\
			 & 1 & 10 & 4 & 1.8 & 4 & 10 & 6.378 & 0.1487 & $4.90(6)\cdot10^{-1}$ & $1.60(8)\cdot10^{-1}$\\
			 & 2 & 20 & 4 & 1.8 & 4 & 10 & 6.416 & 0.1496 & $9.67(9)\cdot10^{-1}$ & $3.0(1)\cdot10^{-1}$\\
			 & 4 & 40 & 4 & 1.8 & 4 & 10 & 6.442 & 0.1502 & $2.02(2)$ & $6.4(2)\cdot10^{-1}$\\
		\hline 
 		\multirow{4}{3.2em}{\text{CGC 10}} 
			 & 1 & 10 & 4 & 1.8 & 4 & 2.5 & 1.598 & 0.03727 & $9.9(2)\cdot10^{-1}$ & $3.3(1)\cdot10^{-1}$\\
			 & 1 & 10 & 4 & 1.8 & 4 & 5 & 3.217 & 0.07502 & $6.83(9)\cdot10^{-1}$ & $2.2(1)\cdot10^{-1}$\\
			 & 1 & 10 & 4 & 1.8 & 4 & 7.5 & 4.799 & 0.1119 & $5.70(5)\cdot10^{-1}$ & $1.79(1)\cdot10^{-1}$\\
			 & 1 & 10 & 4 & 1.8 & 4 & 10 & 6.432 & 0.15 & $4.90(4)\cdot10^{-1}$ & $1.69(6)\cdot10^{-1}$\\
		\hline 
 		\multirow{3}{3.2em}{\text{CGC 11}} 
			 & 0.5 & 20 & 0.18 & 1.8 & 4 & 10 & 0.2801 & 0.001633 & $8(1)\cdot10^{-2}$ & $7(3)\cdot10^{-4}$\\
			 & 1 & 20 & 0.18 & 1.8 & 4 & 10 & 0.2839 & 0.00331 & $1.17(8)\cdot10^{-1}$ & $1.8(2)\cdot10^{-3}$\\
			 & 2 & 20 & 0.18 & 1.8 & 4 & 10 & 0.2844 & 0.006632 & $1.73(4)\cdot10^{-1}$ & $4.4(2)\cdot10^{-3}$\\
		\hline 
 		\multirow{3}{3.2em}{\text{CGC 12}} 
			 & 1 & 10 & 0.18 & 1.8 & 4 & 1.25 & 0.03513 & 0.0008193 & $1.8(1)\cdot10^{-1}$ & $8(1)\cdot10^{-3}$\\
			 & 1 & 10 & 0.18 & 1.8 & 4 & 2.5 & 0.07014 & 0.001635 & $1.20(5)\cdot10^{-1}$ & $3.8(9)\cdot10^{-3}$\\
			 & 1 & 10 & 0.18 & 1.8 & 4 & 5 & 0.1424 & 0.003321 & $1.09(4)\cdot10^{-1}$ & $3.6(2)\cdot10^{-3}$\\
		\hline 
 		\multirow{2}{3.4em}{\text{CGC 13}} 
			 & 1 & 10 & 0.18 & 1.8 & 2.5 & 10 & 0.4213 & 0.009824 & $1.17(2)\cdot10^{-1}$ & $4.2(1)\cdot10^{-3}$\\
			 & 1 & 10 & 0.18 & 1.8 & 5 & 10 & 0.2344 & 0.005465 & $7.2(3)\cdot10^{-2}$ & $1.8(1)\cdot10^{-3}$\\
		\end{tabular}
		\caption{Parameters and simulation results used to produce Figs. \ref{fig:bose} and \ref{fig:EKTvsITA}. All rows are simulated using an initial distribution defined by \Eq{eq:kzdistribution}, $r_0=0.01$. The parameters $\tau_0$, $Q_0$ and $R_0$ are dimensionful, considered here to be in lattice units.}
		\label{tab:scalingCGC}
	\end{table*}
	\begin{table*}[h!]
	\centering
	\begin{tabular}{ c|c|c|c|c|c|c|c|c|c }
		Label & $\tau_0$ & $R_0$ & $Q_0$ & $\xi$ & $\lambda$  & $m_g^2(r_0,\tau_0)$ & $\hat{m}_g^2$ & $v_2^\text{cl.}/\varepsilon_2$ & $v_2^\text{b.e.}/\varepsilon_2$ \\ 
		\hline		\hline 
 		\multirow{3}{3.2em}{\text{TH 1}} 
			 & 0.5 & 20 & 0.69 & 4 & 10 & 0.3532 & 0.002059 & $6.0(9)\cdot10^{-2}$ & $1.5(4)\cdot10^{-3}$\\
			 & 1 & 20 & 0.69 & 4 & 10 & 0.3508 & 0.00409 & $1.14(7)\cdot10^{-1}$ & $2.5(3)\cdot10^{-3}$\\
			 & 2 & 20 & 0.69 & 4 & 10 & 0.3546 & 0.008269 & $1.76(4)\cdot10^{-1}$ & $6.2(2)\cdot10^{-3}$\\
		\hline 
 		\multirow{4}{3.2em}{\text{TH 2}} 
			 & 1 & 10 & 0.69 & 4 & 1.25 & 0.0444 & 0.001035 & $1.72(1)\cdot10^{-1}$ & $1.18(9)\cdot10^{-2}$\\
			 & 1 & 10 & 0.69 & 4 & 2.5 & 0.08881 & 0.002071 & $1.36(4)\cdot10^{-1}$ & $6.9(4)\cdot10^{-3}$\\
			 & 1 & 10 & 0.69 & 4 & 5 & 0.1758 & 0.004101 & $1.10(2)\cdot10^{-1}$ & $4.4(3)\cdot10^{-3}$\\
			 & 1 & 10 & 0.69 & 4 & 10 & 0.3532 & 0.008237 & $8.7(2)\cdot10^{-2}$ & $3.1(1)\cdot10^{-3}$\\
		\hline 
 		\multirow{3}{3.2em}{\text{TH 3}} 
			 & 1 & 10 & 0.69 & 2.5 & 10 & 0.5193 & 0.01211 & $1.15(1)\cdot10^{-1}$ & $5.4(1)\cdot10^{-3}$\\
			 & 1 & 10 & 0.69 & 5 & 10 & 0.2916 & 0.006801 & $7.6(2)\cdot10^{-2}$ & $2.3(2)\cdot10^{-3}$\\
			 & 1 & 10 & 0.69 & 10 & 10 & 0.1554 & 0.003623 & $5.0(3)\cdot10^{-2}$ & $1(2)\cdot10^{-3}$\\
		\hline 
 		\multirow{4}{3.2em}{\text{TH 4}} 
			 & 0.25 & 20 & 2 & 4 & 10 & 3.003 & 0.001028 & $9(3)\cdot10^{-2}$ & $1.5(6)\cdot10^{-3}$\\
			 & 0.5 & 20 & 2 & 4 & 10 & 2.997 & 0.002051 & $2.1(2)\cdot10^{-1}$ & $3(1)\cdot10^{-3}$\\
			 & 1 & 20 & 2 & 4 & 10 & 3.015 & 0.004126 & $3.3(2)\cdot10^{-1}$ & $9(1)\cdot10^{-3}$\\
			 & 2 & 20 & 2 & 4 & 10 & 2.998 & 0.008205 & $5.13(9)\cdot10^{-1}$ & $1.79(4)\cdot10^{-2}$\\
		\hline 
 		\multirow{4}{3.2em}{\text{TH 5}} 
			 & 1 & 10 & 2 & 4 & 1.25 & 0.3768 & 0.001031 & $5.3(1)\cdot10^{-1}$ & $3.4(3)\cdot10^{-2}$\\
			 & 1 & 10 & 2 & 4 & 2.5 & 0.754 & 0.002064 & $4.1(2)\cdot10^{-1}$ & $2.2(1)\cdot10^{-2}$\\
			 & 1 & 10 & 2 & 4 & 5 & 1.492 & 0.004084 & $3.26(7)\cdot10^{-1}$ & $1.41(7)\cdot10^{-2}$\\
			 & 1 & 10 & 2 & 4 & 10 & 2.982 & 0.008163 & $2.47(4)\cdot10^{-1}$ & $8.8(3)\cdot10^{-3}$\\
		\hline 
 		\multirow{3}{3.2em}{\text{TH 6}} 
			 & 1 & 10 & 2 & 2.5 & 10 & 4.486 & 0.01228 & $3.40(3)\cdot10^{-1}$ & $1.61(3)\cdot10^{-2}$\\
			 & 1 & 10 & 2 & 5 & 10 & 2.484 & 0.006799 & $2.33(8)\cdot10^{-1}$ & $6.7(3)\cdot10^{-3}$\\
			 & 1 & 10 & 2 & 10 & 10 & 1.296 & 0.003549 & $1.4(1)\cdot10^{-1}$ & $3.5(5)\cdot10^{-3}$\\
		\hline 
 		\multirow{3}{3.2em}{\text{TH 7}} 
			 & 0.5 & 20 & 1 & 4 & 10 & 0.7498 & 0.002052 & $1.1(1)\cdot10^{-1}$ & $1.2(4)\cdot10^{-3}$\\
			 & 1 & 20 & 1 & 4 & 10 & 0.7539 & 0.004127 & $1.67(7)\cdot10^{-1}$ & $4.0(3)\cdot10^{-3}$\\
			 & 2 & 20 & 1 & 4 & 10 & 0.7503 & 0.008215 & $2.56(6)\cdot10^{-1}$ & $9.1(3)\cdot10^{-3}$\\
		\hline 
 		\multirow{4}{3.2em}{\text{TH 8}} 
			 & 1 & 10 & 1 & 4 & 1.25 & 0.09396 & 0.001029 & $2.4(1)\cdot10^{-1}$ & $1.7(1)\cdot10^{-2}$\\
			 & 1 & 10 & 1 & 4 & 2.5 & 0.1853 & 0.002029 & $2.18(6)\cdot10^{-1}$ & $1.09(8)\cdot10^{-2}$\\
			 & 1 & 10 & 1 & 4 & 5 & 0.3757 & 0.004114 & $1.65(4)\cdot10^{-1}$ & $7.4(3)\cdot10^{-3}$\\
			 & 1 & 10 & 1 & 4 & 10 & 0.7564 & 0.008282 & $1.30(2)\cdot10^{-1}$ & $4.2(2)\cdot10^{-3}$\\
		\hline 
 		\multirow{3}{3.2em}{\text{TH 9}} 
			 & 1 & 10 & 1 & 2.5 & 10 & 1.112 & 0.01218 & $1.69(2)\cdot10^{-1}$ & $7.7(2)\cdot10^{-3}$\\
			 & 1 & 10 & 1 & 5 & 10 & 0.6149 & 0.006733 & $1.08(2)\cdot10^{-1}$ & $3.0(2)\cdot10^{-3}$\\
			 & 1 & 10 & 1 & 10 & 10 & 0.3295 & 0.003608 & $6.8(6)\cdot10^{-2}$ & $1.6(4)\cdot10^{-3}$\\
		\end{tabular}
		\caption{Parameters and simulation results used to produce Figs. \ref{fig:bose} and \ref{fig:EKTvsITA}. All rows are simulated using an initial distribution defined by \Eq{eq:thermaldistribution}, $r_0=0.01$. The parameters $\tau_0$, $Q_0$ and $R_0$ are dimensionful, considered here to be in lattice units.}
		\label{tab:scalingTH}
	\end{table*}
	\begin{table*}[h!]
	\centering
	\begin{tabular}{ c|c|c|c|c|c|c|c|c|c|c }
		Label & $\tau_0$ & $R_0$ & $A$ & $Q_0$ & $\xi$ & $\lambda$  & $m_g^2(r_0,\tau_0)$ & $\hat{m}_g^2$ & $v_2^\text{cl.}/\varepsilon_2$ & $v_2^\text{b.e.}/\varepsilon_2$ \\ 
		\hline		\hline 
 		\multirow{5}{3.2em}{\text{OR}} 
			 & 8 & 20 & 4 & 1.8 & 4 & 10 & 6.497 & 0.606 & $1.703(6)$ & $1.72(1)$\\
			 & 10 & 10 & 4 & 1.8 & 4 & 1 & 0.6383 & 0.1488 & $3.71(1)$ & $5.25(2)$\\
			 & 5 & 10 & 4 & 1.8 & 4 & 2 & 1.281 & 0.1494 & $2.123(9)$ & $2.09(2)$\\
			 & 5 & 10 & 4 & 1.8 & 5 & 10 & 5.266 & 0.614 & $8.15(2)\cdot10^{-1}$ & $7.52(7)\cdot10^{-1}$\\
			 & 10 & 10 & 4 & 1.8 & 10 & 10 & 2.829 & 0.6596 & $7.13(4)\cdot10^{-1}$ & $4.74(5)\cdot10^{-1}$\\
		\end{tabular}
		\caption{Parameters and simulation results used to produce Figs. \ref{fig:bose} and \ref{fig:EKTvsITA}. The values in this table are considered outside of the range of validity of the model, and all rows are simulated using an initial distribution defined by \Eq{eq:kzdistribution}, $r_0=0.01$. The parameters $\tau_0$, $Q_0$ and $R_0$ are dimensionful, considered here to be in lattice units.}
		\label{tab:scalingOR}
	\end{table*}

\section{Initial conditions in nuclear collisions}
\label{sec:initialtables}

 In order to determine realistic values of the scaling variables corresponding to a physical collision systems,  we will re-use the tabulated values of eccentricity $\varepsilon_2$, RMS entropy radius $\sqrt{\left<r^2\right>}$ and entropy density $dS/dy/(\pi \left<r^2\right>)$ (in arbitrary units)  from ref.~\cite{Huss:2020whe}. These initial conditions for PbPb, OO, $p$Pb and pp collision systems
 were generated using the TRENTo initial state model~\cite{Moreland:2014oya, Moreland:2018gsh}. The entropy normalization is not specified, therefore we will use the total entropy per rapidity $dS/dy=11335$ extracted from data for $\sqrt{s_\text{NN}}=2.76\,\text{TeV}$ PbPb 0-10\% collisions~\cite{Hanus:2019fnc}. It is known experimentally that the particle multiplicity ($\propto$ entropy) in nucleus-nucleus collisions scales with the collision energy according to $\propto (\sqrt{s})^{0.31}$ law~\cite{Acharya:2018hhy}. Therefore we use a $(5.02/2.76)^{0.31}$ factor to increase the entropy.
  The next step is to convert this final state entropy to the initial state energy density.  In homogeneous boost-invariant systems the early time non-equilibrium evolution of energy density can be well described by hydrodynamic attractor curves. We will use the following formula derived in ref.~\cite{Giacalone:2019ldn} to relate entropy density per rapidity $(s\tau)_\text{final}$ to initial energy density per rapidity $(e\tau)_0$
\begin{equation}\label{eq:attr}
    (e\tau)_0 = (s\tau)^{3/2}_\text{final} C_\infty^{-\frac{9}{8}}\left(\frac{\eta}{s}\right)^{-1/2}\left( \nu_\text{eff} \frac{256\pi^3}{810}\right)^{-1/2}.
\end{equation}
Here $C_\infty\approx 0.9$ is the property of the hydrodynamic attractor in QCD EKT, $\eta/s$ is the specific shear viscosity and $\nu_\text{eff}\approx 40$ is the effective number of degrees of freedom in the high temperature equilibrium QGP phase. Note that the QCD EKT attractor was obtained for an initial state with only gluons present, so we can identify $(e\tau)_0$ as the energy density of the gluonic degrees of freedom.

 We assume that the averaged initial energy density profile in each centrality class is described by a Gaussian $(e\tau)_0 \exp(-|\xp|^2/R_0^2)$, where $R_0$ is $\sqrt{2/3}$ times the RMS of the entropy profile in ref.~\cite{Huss:2020whe}. Then the energy density at the origin $(e\tau)_0$ is obtained using \Eq{eq:attr} with $\eta/s=0.2$. 
 
 The parameters $(e\tau)_0$ and $R_0$ are not sufficient to determine all the parameters of the initial probability distribution, \Eq{eq:kzdistribution}. Therefore we choose $Q_0=3.0\,\text{GeV}$ in most central 0-10\% PbPb collisions and find that 
 $\hat{A}=A\tau_0/(\xi R_0)\approx 0.00482$ reproduced the estimated initial state energy density $(e\tau)_0$. Keeping the $(e\tau)_0/Q^3_0$ value fixed, we obtain $Q_0$
  for other centralities and collision energies. The results for 5.02~TeV PbPb, OO, $p$Pb and $pp$ collisions are summarized in Tables~\ref{tab:PbPb_input}, \ref{tab:OO_input}, \ref{tab:pPb_input} and \ref{tab:pp_input}. It is straightforward to scale these numbers to correspond to different values of  $\eta/s$.

	\begin{table}[h]
	\centering
	\begin{tabular}{ c|c|c|c|c|c}
		 centrality \% & $R_0$ (fm)& $\varepsilon_2$ & $(e\tau)_0$ GeV${}^3$ & $Q_0$ (GeV) & $\hat{A}$ (GeV) \\ 
		\hline
		 0-10	 & 3.49 & 0.12 & 3.5803 & 3.00 & 0.0048\\
		10-20	 & 3.14 & 0.23 & 2.6581 & 2.72 & 0.0059\\
		20-30	 & 2.86 & 0.30 & 1.9148 & 2.44 & 0.0072\\
		30-40	 & 2.62 & 0.35 & 1.2699 & 2.12 & 0.0091\\
		40-50	 & 2.40 & 0.39 & 0.7658 & 1.79 & 0.0117\\
		50-60	 & 2.21 & 0.41 & 0.4137 & 1.46 & 0.0157\\
		60-70	 & 2.00 & 0.42 & 0.2018 & 1.15 & 0.0220\\
		70-80	 & 1.75 & 0.37 & 0.0846 & 0.86 & 0.0334\\
		80-90	 & 1.48 & 0.27 & 0.0274 & 0.59 & 0.0577\\
		90-100	 & 1.23 & 0.09 & 0.0036 & 0.30 & 0.1367\\
		\end{tabular}
		\caption{Initial state properties for PbPb collision system at $\sqrt{s_\text{NN}}=5.02\,\text{TeV}$ and $\eta/s=0.2$.}
		\label{tab:PbPb_input}
	\end{table}

	\begin{table}[h]
	\centering
	\begin{tabular}{ c|c|c|c|c|c}
		 centrality \% & $R_0$ (fm)& $\varepsilon_2$ & $(e\tau)_0$ GeV${}^3$ & $Q_0$ (GeV) & $\hat{A}$ (GeV) \\ 
		\hline
		 0-10	 & 1.76 & 0.21 & 0.4972 & 1.55 & 0.0185\\
		10-20	 & 1.65 & 0.26 & 0.3470 & 1.38 & 0.0221\\
		20-30	 & 1.55 & 0.29 & 0.2462 & 1.23 & 0.0264\\
		30-40	 & 1.46 & 0.33 & 0.1708 & 1.09 & 0.0318\\
		40-50	 & 1.36 & 0.36 & 0.1168 & 0.96 & 0.0388\\
		50-60	 & 1.24 & 0.38 & 0.0791 & 0.84 & 0.0483\\
		60-70	 & 1.11 & 0.39 & 0.0527 & 0.74 & 0.0618\\
		70-80	 & 0.97 & 0.37 & 0.0349 & 0.64 & 0.0813\\
		80-90	 & 0.84 & 0.32 & 0.0195 & 0.53 & 0.1138\\
		90-100	 & 0.73 & 0.26 & 0.0057 & 0.35 & 0.1962\\
		\end{tabular}
		\caption{Initial state properties for OO collision system at $\sqrt{s_\text{NN}}=5.02\,\text{TeV}$ and $\eta/s=0.2$.}
		\label{tab:OO_input}
	\end{table}

	\begin{table}[h]
	\centering
	\begin{tabular}{ c|c|c|c|c|c}
		 centrality \% & $R_0$ (fm)& $\varepsilon_2$ & $(e\tau)_0$ GeV${}^3$ & $Q_0$ (GeV) & $\hat{A}$ (GeV) \\ 
		\hline
		 0-10	 & 1.17 & 0.34 & 0.2384 & 1.22 & 0.0355\\
		10-20	 & 1.16 & 0.34 & 0.1625 & 1.07 & 0.0408\\
		20-30	 & 1.15 & 0.35 & 0.1236 & 0.98 & 0.0448\\
		30-40	 & 1.13 & 0.34 & 0.1017 & 0.92 & 0.0488\\
		40-50	 & 1.10 & 0.35 & 0.0829 & 0.86 & 0.0536\\
		50-60	 & 1.05 & 0.34 & 0.0675 & 0.80 & 0.0602\\
		60-70	 & 1.01 & 0.36 & 0.0489 & 0.72 & 0.0694\\
		70-80	 & 0.94 & 0.35 & 0.0357 & 0.65 & 0.0833\\
		80-90	 & 0.86 & 0.33 & 0.0222 & 0.55 & 0.1068\\
		90-100	 & 0.75 & 0.28 & 0.0076 & 0.39 & 0.1740\\
		\end{tabular}
		\caption{Initial state properties for $p$Pb collision system at $\sqrt{s_\text{NN}}=5.02\,\text{TeV}$ and $\eta/s=0.2$.}
		\label{tab:pPb_input}
	\end{table}

	\begin{table}[h]
	\centering
	\begin{tabular}{ c|c|c|c|c|c}
		 centrality \% & $R_0$ (fm)& $\varepsilon_2$ & $(e\tau)_0$ GeV${}^3$ & $Q_0$ (GeV) & $\hat{A}$ (GeV) \\ 
		\hline
		 0-10	 & 0.88 & 0.33 & 0.1023 & 0.92 & 0.0624\\
		10-20	 & 0.89 & 0.33 & 0.0644 & 0.79 & 0.0724\\
		20-30	 & 0.88 & 0.32 & 0.0503 & 0.72 & 0.0789\\
		30-40	 & 0.88 & 0.33 & 0.0404 & 0.67 & 0.0849\\
		40-50	 & 0.86 & 0.32 & 0.0345 & 0.64 & 0.0914\\
		50-60	 & 0.87 & 0.32 & 0.0270 & 0.59 & 0.0991\\
		60-70	 & 0.84 & 0.31 & 0.0219 & 0.55 & 0.1090\\
		70-80	 & 0.83 & 0.32 & 0.0162 & 0.50 & 0.1231\\
		80-90	 & 0.80 & 0.31 & 0.0105 & 0.43 & 0.1463\\
		90-100	 & 0.76 & 0.28 & 0.0042 & 0.32 & 0.2111\\
		\end{tabular}
		\caption{Initial state properties for $pp$ collision system at $\sqrt{s_\text{NN}}=5.02\,\text{TeV}$ and $\eta/s=0.2$.}
		\label{tab:pp_input}
	\end{table}

\end{document}